\documentclass[aps,pre,floatfix,twocolumn,nofootinbib]{revtex4-2}

\usepackage{lineno}

\usepackage{bm}
\usepackage[dvips]{epsfig}
\usepackage{graphicx} 
\usepackage{svg}
\usepackage{subcaption}
\usepackage{amsmath} 
\usepackage{amssymb}
\usepackage{appendix}
\usepackage{xcolor}
\usepackage{listings}
\usepackage{url}
\usepackage{tabularx}
\usepackage{array}
\usepackage{ragged2e}
\usepackage{soul}
\usepackage{bbm}
\usepackage{ulem}

\renewcommand{\st}[1]{\phantom{#1}}

\newcommand{\nn}{\nonumber \\}
\renewcommand{\d}{{\rm d}}
\newcommand{\bro}{\boldsymbol{\rho}}
\newcommand{\br}{\textbf{r}}
\newcommand{\bs}{\textbf{s}}
\newcommand{\bq}{\textbf{q}}
\newcommand{\bQ}{\textbf{Q}}
\newcommand{\bS}{\textbf{S}}
\newcommand{\bp}{\textbf{p}}

\newcommand{\qE}{\mathsf{E}}

\newcommand{\bu}{\mathbf u}
\newcommand{\bv}{\mathbf v}

\begin{document}
\title{Propagation of Two-Photon Zernike States in Atmospheric Turbulence}

\author{Hakob Avetisyan} 
\affiliation{
Alikhanyan National Laboratory (Yerevan Physics Institute), 2 Alikhanyan Brothers Street, Yerevan 0036, Armenia} 
\email{h.avetisyan@yerphi.am}
\author{Vahagn Abgaryan} 
\affiliation{Alikhanyan National Laboratory (Yerevan Physics Institute), 2 Alikhanyan Brothers Street, Yerevan 0036, Armenia}
\affiliation{Joint Institute for Nuclear Research, 6 Joliot-Curie St.,
Dubna, 141980,  Russia}
\email{vahagnab@googlemail.com}

\begin{abstract}
We analyze propagation and detection of two-photon states expanded in Zernike modes through atmospheric turbulence using the extended Huygens–Fresnel formalism. For SPDC states prepared with a single Zernike pump mode, 
we analytically reduce the 8-dimensional continuous propagation integrals to an exact, discrete modal expansion.   
In the absence of turbulence, Zernike addition enforces conservation of azimuthal index and a strict radial-order bound. Turbulence relaxes these constraints, driving structured azimuthal and radial crosstalk dominated by low-order aberration modes. 
\st{By explicitly removing the lowest-order terms from the discrete turbulence sum, we demonstrate that partial adaptive optics correcting only up to the sixth radial order is sufficient to heavily suppress this crosstalk and restore near-ideal spatial correlations.}
\end{abstract}


\maketitle

\section{Introduction}
Spatially structured optical fields provide a high-dimensional resource for quantum optics and quantum information processing, enabling enhanced encoding capacity, robustness, and sensitivity in tasks ranging from quantum communication to quantum sensing \cite{Allen, Bazhenov, Beijersbergen, Padgett, Agarwal, Mair, Vaziri, Malik}. 
In particular, the transverse spatial degrees of freedom of photons generated via spontaneous parametric down-conversion (SPDC) can exhibit strong correlations and entanglement, which have been extensively studied using Hermite–Gaussian (HG) and Laguerre–Gaussian (LG) mode bases \cite{walborn04, walborn05}. These modal representations have played a central role in establishing conservation laws, selection rules, and entanglement properties of quantum states of spatially structured light. 

In practical free-space implementations, however, spatially encoded quantum states must propagate through random or turbulent media, most notably the atmosphere. Atmospheric turbulence introduces random phase distortions that couple transverse modes, degrade modal orthogonality, and weaken spatial correlations.
This manifests as severe intermodal crosstalk—most notably observed as the degradation of orbital angular momentum (OAM) selection rules—
effectively acting as a noisy quantum channel \cite{Andrews, Boyd:11, Ren:13, krenn, vogel, Pirandola2022, PhysRevLett.94.153901, Tyler:09}. Understanding how turbulence affects different spatial encodings is therefore a central problem in the development of robust quantum photonic technologies.

Turbulence-induced degradation of spatial correlations has been extensively investigated in HG and LG modal bases, frequently, within the extended Huygens–Fresnel formalism \cite{avetisyan16, avetisyan17}. However, both, the robustness of correlations and the structure of intermodal coupling depend strongly on the chosen representation. In this work, we retain the same propagation framework while reformulating the analysis in the Zernike basis.


Motivated by this observation, we recently introduced Zernike polynomials as a physically meaningful basis for describing quantum states of light \cite{Avetisyan25Z}.  
They are widely used in classical optics to represent wavefront aberrations, with low-order polynomials corresponding directly to familiar distortions such as piston, tilt, defocus, astigmatism, coma, and trefoil \cite{Noll, bezdidko, Fried65}. In the quantum context, we showed that SPDC can generate entanglement in the Zernike basis and that the associated expansion coefficients obey exact selection rules closely analogous to angular-momentum conservation in LG modes, while simultaneously encoding radial-order constraints absent in conventional OAM descriptions.

The direct correspondence between Zernike modes and classical aberrations suggests that this basis may offer distinct advantages for analyzing and mitigating turbulence-induced decoherence. Since atmospheric turbulence is itself commonly described in terms of low-order aberrations \cite{Andrews, Noll}, one may expect that its dominant effects are naturally captured by a restricted subset of Zernike modes. 

In this work, we develop a theoretical framework for the propagation and detection of Zernike-mode two-photon states through atmospheric turbulence. Using the extended Huygens–Fresnel principle, we incorporate random phase distortions into the Zernike-mode representation of the optical field and derive analytical expressions for the joint two-photon detection probability in the far field. The formalism allows us to separate the roles of state preparation, modal structure, and turbulence-induced mode coupling in a transparent and systematic manner.

Applying the theory to entangled photon pairs generated via SPDC, we analyze how turbulence modifies the exact selection rules that hold in the absence of random media. We show that while turbulence relaxes these constraints, the resulting degradation of spatial correlations is dominated by a small set of low-order Zernike modes associated with classical aberrations. This hierarchical coupling behavior contrasts with the extensive mode coupling observed in LG- and HG-based descriptions \cite{PhysRevLett.94.153901, Tyler:09, avetisyan17}, and provides a clear physical interpretation of turbulence-induced decoherence in terms of aberration dynamics.

In the present work, we adopt the same propagation framework but reformulate it in the Zernike basis, which is naturally adapted to aberration-driven distortions and allows the turbulence-induced coupling structure to be analyzed in a physically transparent manner.
In addition to their application to quantum field propagation in turbulence, the Zernike and Fourier--Zernike coupling identities derived here provide a closed algebraic framework that may be useful in other contexts involving aberration-driven mode coupling.

The paper is organized as follows. Section \ref{formalism} introduces the Zernike-mode formalism used throughout the work. In Section \ref{2a} summarizes the properties of Zernike polynomials, their Fourier transforms, and the associated algebraic identities (some of which are new, to the best of our knowledge) required for two-photon mode coupling. Section \ref{2b} and \ref{2c} incorporate atmospheric turbulence into the Zernike-mode representation of the optical field operators using the extended Huygens–Fresnel principle. In Section \ref{sec3}, we apply the formalism to the propagation and detection of entangled photon pairs generated via SPDC and derive analytical expressions for joint two-photon detection probabilities in atmospheric turbulence channel. Section \ref{discuss} discusses the physical implications of the results. Section \ref{summary} summarizes the main conclusions and outlines directions for future work.

\section{Zernike-mode formalism and field propagation through turbulence.}\label{formalism}
\subsection{Zernike modes and mode-coupling identities.}\label{2a}
\st{This subsection summarizes the Zernike and Fourier–Zernike identities required for subsequent derivations.}
The foundational definitions of the quantum Zernike basis and free-space propagation operators were introduced in our recent work \cite{Avetisyan25Z}. We briefly summarize the essential Zernike and Fourier-Zernike identities here to establish notation before introducing the novel turbulent propagation framework.
Zernike modes form a complete orthonormal basis on the unit disk and provide a natural representation for optical fields defined on a circular pupil.

Unlike LG modes, Zernike modes impose explicit radial-order constraints. They are defined for integers $m$, $n$ with $n - |m|\geq 0$ and even as
\begin{align}
    Z_n^m(\bro) &= 
    \begin{cases}
        \sqrt{n+1}\,R_n^{|m|}(\rho)e^{im\theta}, &\quad 0\leq \rho \leq 1, \\ 
        \qquad 0 & \quad \rho>1,
    \end{cases}
\end{align}
where $\bro=(\rho,\theta), \,\, 0\leq \theta<2\pi,$ 
are the polar coordinates $(\rho = \sqrt{x^2+y^2}, \quad \theta = \arctan(y/x))$,
and the radial polynomials $R^{|m|}_n$ are given by
\begin{align}
    R_n^{|m|}(\rho,\theta) = \sum_{k=0}^{\frac{n-|m|}{2}}\frac{(-1)^k(n-k)!}{k!\left(\frac{n+m}{2}-k\right)!\left(\frac{n-m}{2}-k\right)!}\rho^{n-2k}.
\end{align}
We adopt the Fourier-transform convention
\begin{align}
    \widetilde{f}(\textbf{k}) &= \int \d^2\rho \, f(\br)\,e^{2\pi i\, \br\cdot \textbf{k}},
\end{align}
under which the Fourier transforms of Zernike modes exhibit a simple Bessel representation \cite{Noll}
\begin{align} 
    \widetilde{Z}_n^m(\bq)
    &=\int_0^1 \d \rho\,\rho\int_0^{2\pi}
    \d\theta\,Z_n^m(\rho,\theta)e^{2\pi i \rho q \cos(\theta-\phi)}
   \nn&=
   2\pi i^n\sqrt{n+1}\frac{J_{n+1}(2\pi q)}{2\pi q} e^{im\phi},
  \label{Z-fourier}
  \end{align}
where $\bq = (q,\phi)$ are polar coordinates in Fourier space.

The Zernike modes and their Fourier counterparts satisfy orthogonality and completeness relations. In real space (pupil plane), they are
\begin{align}
    \int \d^2 s\, Z_n^{m}(\mathbf s)\,Z_{n'}^{m'*}(\mathbf s)  &= \pi\,\delta_{n n'}\,\delta_{m m'},\label{orth}\\
    \sum_{n,m} Z_n^{m}(\mathbf s_1)\,Z_n^{m*}(\mathbf s)  &=\pi \,\delta_{D}(\mathbf s_1-\mathbf s).\label{compl}
\end{align}
where $\delta_{D}(\bs_1-\bs)=\mathbbm{1}_{D}(\mathbf{s}_{1})\mathbbm{1}_{D}(\mathbf{s}_{2})\,\delta(\bs_{1}-\bs_{2})$ is the delta function defined as the kernel of the projection onto the functions supported on a subset of the unit disk $D$. $\mathbbm{1}_{D}(\bs)$ denotes the indicator function of $D$, i.e., $\mathbbm{1}_{D}(\bs)=1$ for $\bs \in D$, $\mathbbm{1}_{D}(\bs)=0$ otherwise.
The analogous expressions for the Fourier space (image plane) are 
\begin{align}
     \int \d^2 q\, \widetilde{Z}_n^{m}(\bq)\,\widetilde{Z}_{n'}^{m'*}(\bq)  &= \pi\,\delta_{n n'}\,\delta_{m m'},\label{F_orth11}\\
     \sum_{n,m} \widetilde{Z}_n^{m}(\bq_1)\,\widetilde{Z}_n^{m*}(\bq)  &=\pi \, \frac{J_{1}(2\pi|\bq-\bq_{1}|)}{|\bq-\bq_1|}
     \nn&=\pi\,\widetilde{Z}_{0}^{0}(\bq-\bq_{1})\label{F_compl11}.
\end{align}
The right-hand side of Eq.~(\ref{F_compl11})   is the  integral kernel for the projector onto the subspace of $\mathcal{F}\left\{L^{2}_{D}\right\}\subset L^{2}_{\mathbb{R}^{2}}$ i.e. functions with Fourier pre-image supported on the unit disk. Furthermore, the natural action of the projector on its eigenspace is guaranteed by the obvious relation 
\begin{align}
\int \d^2 q_{1}  \widetilde{Z}_{0}^{0}(\bq-\bq_{1}) \widetilde{Z}_{n}^{m}(\bq_1)=\pi \,\widetilde{Z}_{n}^{m}(\bq).
\end{align} 
These properties ensure a complete modal description of pupil- and image-plane optical fields. All projector identities below (including the kernels $\delta_D$ and $\widetilde{Z}_0^0$) are to be understood only under integration against admissible test functions in the corresponding subspace.

Mode coupling in the Zernike basis is governed by two complementary sets of coefficients. The first are the $A-$coefficients, defined by the triple overlap integral \cite{tango}
\begin{align}
    A_{n_1 n_2 N}^{m_1 m_2 M}
    = \frac{1}{\pi}\int d^{2}s\;
    Z_{n_1}^{m_1}(\bs)\,Z_{n_2}^{m_2}(\bs)\,Z_{N}^{M*}(\bs),\label{A_def}
\end{align}
which encode exact Zernike-mode addition rules via Clebsch-Gordan (CG) coefficients (see Appendix \ref{app:A_defs}). This is equivalent to
\begin{align}
    Z_{n_1}^{m_1}(\bs)\,Z_{n_2}^{m_2}(\bs)=\sum_{NM}A_{n_1 n_2 N}^{m_1 m_2 M}\,Z_{N}^{M}(\bs)\label{A_cg}.
\end{align}
While the foundational properties of quantum Zernike modes were established in Ref. \cite{Avetisyan25Z}, evaluating turbulence integrals requires a specialized set of Fourier-domain overlap identities. To the best of our knowledge, the following summation identities (Eqs. \ref{A_Z_proj}-\ref{F_A_Z_proj}, \ref{eq:F_G_Z_proj}-\ref{GA_orth}) are newly derived here to enable the discrete tensor collapse of the turbulent propagation integrals (see Appendix \ref{ap:identities} for derivations). From the definition \eqref{A_def} there follow additional identities
\begin{align}
    \sum_{\substack{n_1,m_1\\n_2,m_2}}
    A_{n_1 n_2 N}^{m_1 m_2 M}\; &
    Z_{n_1}^{m_1}(\mathbf s_1)\,Z_{n_2}^{m_2}(\mathbf s_2)
    \nn&= 
    \pi\,
    Z_N^M(\mathbf s_1)\,\delta_{D}(\mathbf s_1-\mathbf s_2),\label{A_Z_proj}
\end{align}
\begin{align}
    \sum_{\substack{n_1,m_1\\n_2,m_2}}
    A_{n_1 n_2 N}^{m_1 m_2 M}\;&
    \widetilde{Z}_{n_1}^{m_1}(\bq_1)\,
    \widetilde{Z}_{n_2}^{m_2}(\bq_2)
    \nn&=\pi\,\,
    \widetilde{Z}_N^M(\bq_1+\bq_2). \label{F_A_Z_proj}
\end{align}

A second set of coefficients, denoted $\Gamma$, arises from convolution and linearization of Zernike modes. The corresponding real-space identity, for $0\leq  |\bs|\leq 2$, \cite{kintner, Janssen2011}
\begin{align}
    \left(Z_n^m \ast Z_{n'}^{m'}\right)(\bs) &= \sum_{n''m''}\Gamma_{nn'n''}^{mm'm''} Z_{n''}^{m''}(\bs/2), \label{eq:kintner_janssen}
\end{align}
and its obvious implication, the linearization of Fourier-Zernike modes,
\begin{align}
    \widetilde{Z}_n^m(\bq) \widetilde{Z}_{n'}^{m'}(\bq) &= 4\sum_{n''m''}\Gamma_{nn'n''}^{m
    m'm''} \widetilde{Z}_{n''}^{m''}(2\bq),\label{eq:F_kintner_janssen}
\end{align}
which follows from applying the convolution theorem to Eq.~\eqref{eq:kintner_janssen}. (notice, Eq.~\eqref{eq:kintner_janssen} slightly differs from that of Refs. \cite{kintner, Janssen2011} – the second function in the convolution is not conjugated here.) Hence, the definition of the coefficients $\Gamma$ can be taken as 
\begin{align}
    \Gamma_{nn'n''}^{m
    m'm''}  = \frac{1}{\pi}\int \d^2q \widetilde{Z}_n^m(\bq) \widetilde{Z}_{n'}^{m'}(\bq)\widetilde{Z}_{n''}^{m''*}(2\bq).\label{eq:Gamma_def}
\end{align}
From \eqref{eq:Gamma_def} also follows (see Appendix \ref{ap:identities})
\begin{align}
     \sum_{\substack{n_1,m_1\\n_2,m_2}} &\Gamma_{n_1n_2N}^{m_1m_2M} \widetilde{Z}_{n_1}^{m_1*}(\bq_1) \widetilde{Z}_{n_2}^{m_2*}(\bq_2) 
     \nn&=
     \pi\int \d^2q \,
     \widetilde{Z}_{N}^{M*}(2\bq)\,\widetilde{Z}_0^0(\bq-\bq_1)\,\widetilde{Z}_0^0(\bq-\bq_2)
     ,\label{eq:F_G_Z_proj}
\end{align}
and double inverse Fourier transforming, we get
\begin{align}
     \sum_{\substack{n_1,m_1\\n_2,m_2}} \Gamma_{n_1n_2N}^{m_1m_2M}& Z_{n_1}^{m_1*}(\bs_1) Z_{n_2}^{m_2*}(\bs_2) 
     \nn&=
     \frac{\pi}{4}Z_N^{M*}\!\left(\frac{\bs_1+\bs_2}{2}\right)\mathbbm{1}_{D}(\mathbf{s}_{1})\mathbbm{1}_{D}(\mathbf{s}_{2}).\label{eq:GZZsum}
\end{align}
\eqref{eq:F_G_Z_proj} and \eqref{eq:GZZsum} are the counterparts of \eqref{A_Z_proj} and \eqref{F_A_Z_proj}: 
the $\Gamma-$coefficients play a complementary role.
Another important identity is 
\begin{align}
    \sum_{nm}Z_n^m(\bs)\widetilde{Z}_n^{m*}(\bq) &= \pi\,\mathbbm{1}_D(\bs)\, e^{-2\pi i \bs\cdot\bq},\label{Z_F_Z_compl}
\end{align}
from which the following relation follows immediately
\begin{align}
     \sum_{\substack{n_1,m_1\\n_2,m_2}} \Gamma_{n_1n_2N}^{m_1m_2M*} A_{n_1n_2N_1}^{m_1m_2M_1}  =
     \frac{\pi}{4}\,\delta_{NN_1}\,\delta_{MM_1},
     \label{GA_orth}
\end{align}
and similarly under permutations of $((n_1,m_1),$ $(n_2,m_2),$ and $(N,M))$.

The derivations of the above identities and contraction identities involving the $A-$ and $\Gamma-$coefficients, as well as their explicit representations, are collected in Appendix \ref{ap:identities}. \st{To the best of our knowledge, several identities derived here do not appear in the existing literature.}

We note that identities derived above (e.g. ~\eqref{GA_orth}) rely only on completeness and orthogonality of the underlying mode set, and therefore have analogs in other orthonormal bases. The distinctive features of the Zernike representation arise from the specific algebraic structure of the associated coupling coefficients and their radial and azimuthal selection rules. 

The $A-$coefficients enforce real-space coincidence and Fourier-domain mode addition, while the $\Gamma-$coefficients provide the complementary coupling structure connecting real and Fourier representations that will later be shown to govern non-collinear  and collinear two-photon correlations, respectively.

\subsection{Zernike expansion of the pupil function and field representation.}\label{2b}
To connect the Zernike-mode formalism with physical optical fields, we represent the generalized pupil function in the Zernike basis. For a circular pupil of radius $R$, the complex pupil function can be written as
\begin{align}
    P(R\rho, \theta) &= A(R\rho, \theta) \exp[i\Phi(R\rho, \theta)] 
    \nn&=
    \sum_{n=0}^\infty \sum_{m=-n}^n b_{mn} Z_n^m(\rho, \theta),
    \label{pupil_exp}
\end{align}
where the complex coefficients $b_{nm}$ account for both phase aberrations and amplitude variations across the pupil. In the special case of a purely phase-aberrated wavefront, this reduces to expansion frequently used in optics  
\begin{align}
    \exp[i\Phi(R\rho, \theta)] =
    \sum_{n=0}^\infty \sum_{m=-n}^n a_{mn} Z_n^m(\rho, \theta).
    \label{phase_exp}
\end{align}
with real-valued coefficients $a_{nm}$ directly associated with specific aberration modes. While the $a_{nm}$ coefficients admit a direct physical interpretation in terms of classical aberrations, the more general $b_{nm}$ representation provides a compact and flexible description of realistic optical fields, in which only a limited number of low-order modes typically contribute significantly. A key consequence is that, within the Fraunhofer approximation, a field expanded in Zernike modes in the pupil plane maps to an image-plane field obtained by replacing each Zernike polynomial by its Fourier–Zernike transform with unchanged expansion coefficients, a property that will be exploited in the following analysis.

\subsection{Field operators and extended Huygens–Fresnel principle.}\label{2c}
Having established the free-space Zernike operators \cite{Avetisyan25Z}, we now introduce the central theoretical contribution of this work: incorporating atmospheric phase distortions into the quantum Zernike framework using the extended Huygens-Fresnel principle.
Consider the propagation of optical fields expanded in the Zernike basis through a turbulent medium. Starting from a pupil-plane representation of the field in terms of Zernike modes, we incorporate the effects of atmospheric turbulence using the extended Huygens–Fresnel principle, which provides a statistical description of wave propagation in random media. In this framework, turbulence enters as a stochastic phase perturbation accumulated along the propagation path, allowing ensemble-averaged field correlations to be evaluated analytically. This approach enables a systematic treatment of how Zernike-mode components are modified under propagation while remaining naturally adapted to rotationally symmetric, statistically isotropic turbulence.

Within the extended Huygens–Fresnel formalism, the field propagated from the source plane to a distance $z$ at transverse coordinate $\bro=(x,y)$ is expressed as a superposition of secondary wavelets, each acquiring a random, turbulence-induced phase shift
\begin{align}\label{ext_Huygens_Fresnel} 
    U(\bro,z)=\frac{ke^{ikz}}{2\pi iz}
    \int_{D}\d^2s\ U_0(\bs,0)
    e^{\frac{ik\vert\bs-\bro\vert^2}{2z}+\psi(\bro,\bs)},
\end{align}
where $D$ is the pupil domain. In addition to the deterministic Fresnel phase, propagation through the random medium introduces a stochastic contribution $\psi(\bro,\bs)$, which accounts for the cumulative effect of refractive-index fluctuations along the propagation path. This random phase is treated as a statistically homogeneous and isotropic process --a standard and widely accepted assumption within the Kolmogorov model of atmospheric turbulence \cite{Andrews}-- so that ensemble-averaged field correlations can be evaluated in terms of the spatial separation of propagation paths rather than individual realizations of the medium. $\psi(\bro,\bs)=\psi_1(\bro, \bs)+\psi_2(\bro, \bs)$ where $\psi_1(\bro, \bs)$ and $\psi_2(\bro, \bs)$ are first- and second-order perturbations, respectively.

If, in addition to the Fresnel approximation, the stronger Fraunhofer approximation ($z> ks^2_{\text{max}}/\pi=4kD/\pi^2$) is satisfied, then the field is determined as
\begin{align}\label{ext_Huygens_Fraunhofer}
U(\bro,z)=\frac{ke^{ikz}e^{\frac{ik}{2z}\rho^2}}{2\pi iz}\!
\int_{D}\d^2s\, U_0(\bs,0)
e^{-\frac{ik}{z}\bro\cdot\bs+\psi(\bro,\bs)}.
\end{align}
The Fraunhofer representation is particularly simple because a pupil-plane function expressed in Zernike modes $Z^m_n(\bro)$ has an image-plane counterpart obtained simply by substituting their Fourier transforms $\widetilde{Z}^m_n(\bq)$ for $Z^m_n(\bro)$ \cite{kintner1}. This structure is expected to remain largely intact once weak atmospheric turbulence is incorporated.

Having established the free-space Zernike operators \cite{Avetisyan25Z}, we now introduce the central theoretical contribution of this work: incorporating atmospheric phase distortions into the quantum Zernike framework. To achieve this, we reformulate the extended Huygens-Fresnel description in terms of field operators. As derived in our previous work \cite{Avetisyan25Z}, the positive-frequency field operator at position $\br=(\bro,z)$ in the paraxial approximation takes the Zernike-mode form 
\begin{align}\label{QEFresnel}
    \qE^{(+)}(\br) &=
    \frac{ke^{ikz}}{2iz\pi^\frac{3}{2}}\sum_{mn} \widetilde{z}_{n}^{m} 
    \int \d^2 s \,Z_{n}^{m}(\bs) e^{\frac{ik}{2z}|\bro-\bs|^2},
\end{align}
where $\widetilde{z}_{n}^{m}=\frac{1}{\sqrt{\pi}}\int\d^2q \widetilde{Z}_n^m(\bq)a(\bq)$ are the Zernike-mode annihilation operators, acting on the vacuum state as $|z_n^m\rangle = (\widetilde{z}_n^m)^\dagger\!|0\rangle$ and satisfying the discrete bosonic commutation relations $[\widetilde{z}_n^m, \widetilde{z}_{n'}^{m'\dagger}]=\delta_{mm'}\delta_{nn'}$.
In the quantum description, the classical Zernike expansion coefficients $b_{nm}$ in \eqref{pupil_exp} are promoted to annihilation operators $\widetilde{z}_n^m$, preserving the Zernike-mode structure of the field while enabling a direct operator-based treatment of propagation and detection.

In the Fraunhofer zone the field operator assumes the form,
\begin{align}
    \qE^{(+)}(\br) &=\frac{k e^{ikz+\frac{ik\rho^2}{2z}}}{iz\sqrt{\pi}}\sum_{mn}\widetilde{Z}_n^m(\bro)\, \widetilde{z}_{n}^{m},
    \label{field-op-1pZFrhf}
\end{align}
where $\bro = 2\pi\,\bq \, z/k$ is the position at the detection plane corresponding to the Fourier component $\bq$. 
For single-photon or collinear two-photon detection scenarios, factors 
$e^{ikz}$ and $e^{\frac{ik\rho^2}{2z}}$
are irrelevant. Only for the joint two-photon detection case, they must either be retained or assumed to be removed by adaptive optics. 
To account for the effect of the medium on the Fraunhofer pattern, one can add the complex random phase $\psi$ to the propagator in \eqref{Z-fourier}
\begin{align}     
    \qE^{(+)}(\br) =\frac{ke^{ikz}}{2iz \pi^\frac{3}{2}}\sum_{mn}
    \widetilde{z}_{n}^{m}
    \!\int\! \d^2 s  Z_n^m(\bs)
    e^{i \frac{k}{z}\, \bs \cdot\bro+\psi(\bro,\bs)}.\label{qE-Fhfr-turb}
\end{align}

\section{Propagation and detection of two-photon Zernike states}\label{sec3}
We now apply the propagation formalism developed in Section \ref{formalism} to spatially entangled degenerate ($\omega_1=\omega_2$) photon pairs generated via spontaneous parametric down-conversion in the paraxial, monochromatic approximations \cite{walborn}. The physical configuration (illustrated in Figure \ref{fig:setup}) consists of an SPDC crystal of thickness $L$ pumped by a structured beam. This generates entangled photon pairs that subsequently propagate through a turbulent atmospheric channel of finite length $z$ before reaching the far-field detection plane:
\begin{figure}
    \centering
    \includegraphics[width=0.9\linewidth]
    {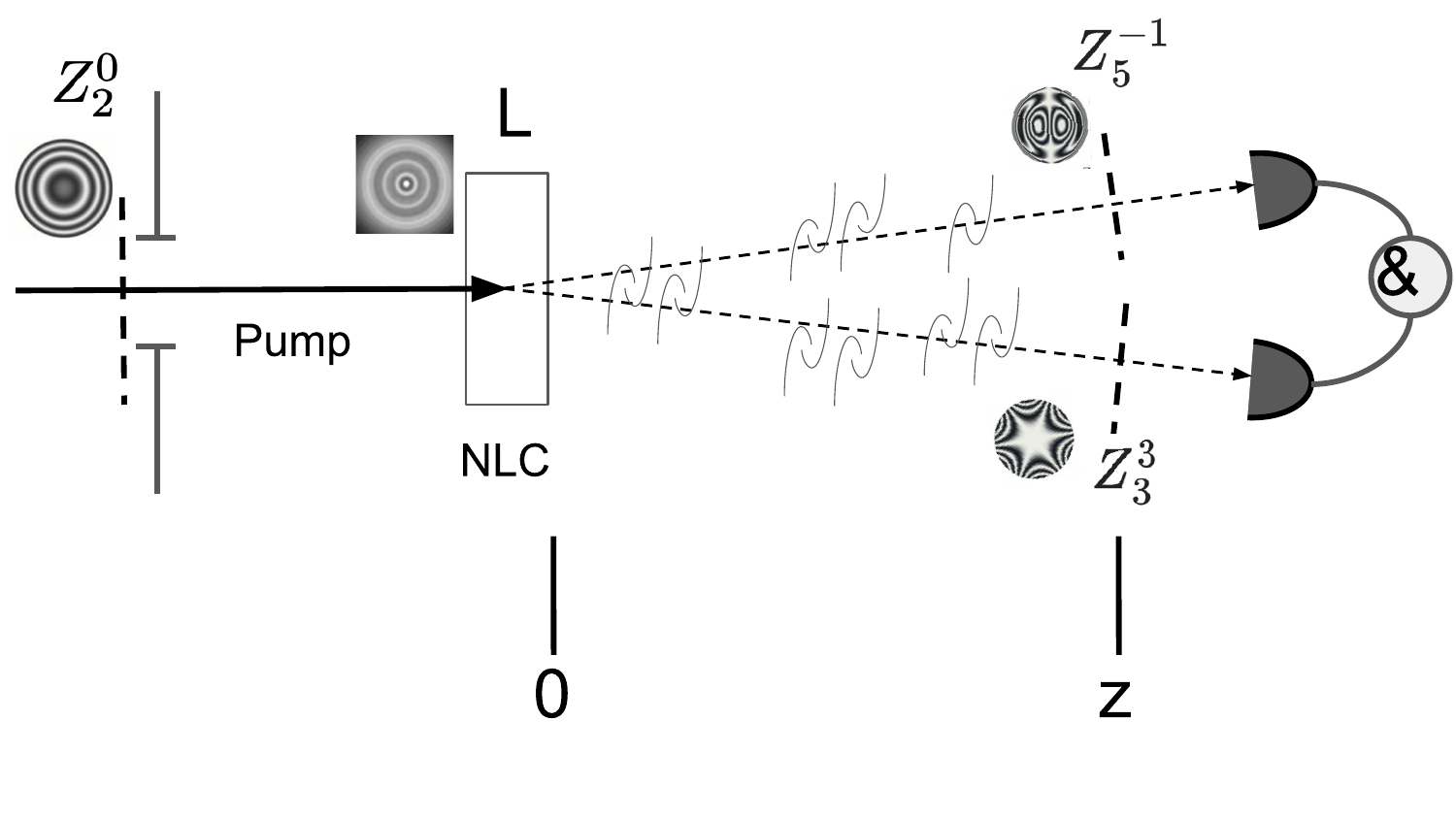}
    \caption{\justifying Schematic of the physical configuration. A structured pump beam prepared in a specific Zernike mode (e.g., $Z_2^0$) undergoes spontaneous parametric down-conversion in a thin nonlinear crystal (NLC) of thickness $L$ at $z=0$. The generated spatially entangled photon pairs propagate through a turbulent atmospheric channel over a distance $z$. At the far-field detection plane at $z$, the photons are projected onto target spatial Zernike modes (e.g., $Z_5^{-1}$ and $Z_3^3$) and their joint detection probability is recorded via a coincidence counter.}
    \label{fig:setup}
\end{figure}
\begin{align} \label{spdc_thin}
\vert\psi\rangle &= \iint \d^2q_1 \d^2q_2 
\widetilde{E}_p\left(\bq_1+\bq_2 \right)
\nn&\qquad\times
\text{sinc}\!\left(\frac{L(|\bq_1-\bq_2|^2)}{4K}\right)
\hat{a}^{\dagger}(\bq_1)
\hat{a}^{\dagger}(\bq_2)\vert 0\rangle,
\end{align} 
where $\bq_1$ and $\bq_2$ are the transverse components of the down-converted wave vectors, $\widetilde{E}_p(\bq)$ is the angular spectrum of the pump-beam, $k_p$ is the wave number of the pump-beam, $L$ the crystal thickness. 
The state (\ref{spdc_thin}) written in the Zernike mode basis has the form \cite{Avetisyan25Z}
\begin{align}
    |\psi\rangle &=     \sum_{\substack{m_1m_2\\n_1n_2}} \zeta_{n_1n_2}^{m_1m_2} |z_{n_1}^{m_1},z_{n_2}^{m_2}\rangle,\label{tpz}
\end{align}
where $\zeta_{n_1n_2}^{m_1m_2}=\langle z_{n_1}^{m_1},z_{n_2}^{m_2}|\psi_2\rangle$.
The two-photon state in \eqref{tpz} provides the setting in which the $A-$ and $\Gamma-$structures discussed in Section \ref{formalism} naturally enter the two-photon problem. In particular, interesting subclasses arise when the coefficients 
$\zeta_{n_1n_2}^{m_1m_2}$ acquire the structured forms associated with the $A-$ and $\Gamma-$coefficients, leading to the following two-photon states:
\begin{align}
    |\psi&\rangle_A =     \sum_{\substack{m_1m_2\\n_1n_2}} A_{n_1n_2N}^{m_1m_2M} |z_{n_1}^{m_1},z_{n_2}^{m_2}\rangle 
    \nn&=
    \iint\d^2q_1\,\d^2q_2\, \widetilde{Z}_{N}^{M}(\bq_1+\bq_2) a^\dagger(\bq_1) a^\dagger(\bq_2)|0\rangle
    ,\label{tpz_A_q}
\end{align}
which corresponds to the thin-crystal approximation of SPDC, with $\widetilde{E}_p\left(\bq_1+\bq_2 \right) = \widetilde{Z}_N^M\left(\bq_1+\bq_2 \right)$,
and 
\begin{align}
    |\psi\rangle_{\Gamma} &=     \sum_{\substack{m_1m_2\\n_1n_2}} \Gamma_{n_1n_2N}^{m_1m_2M} |z_{n_1}^{m_1},z_{n_2}^{m_2}\rangle \label{tpz_Gamma}
    \\&= 
    \int\d^2q \widetilde{Z}_N^M(2\bq) \left(\hat{\zeta}^\dagger(\bq)\right)^2|0\rangle
    ,\label{tpz_Gamma_q}
\end{align}
where \eqref{F_A_Z_proj} and \eqref{eq:F_G_Z_proj} were used, and $\hat{\zeta}^\dagger(\bq)\equiv \int\d^2q'\widetilde{Z}_0^0(\bq-\bq')a^\dagger(\bq')$. This state is a superposition of two-photon states where both photons have passed through an aperture $Z_0^0$, with pump envelope $\widetilde{Z}_N^M(2\bq)$.
Note that, for a pump with $N=M=0$, the state \eqref{tpz_Gamma} is separable in the Zernike-mode basis. This does not contradict the non-separable structure of the transverse-momentum representation \eqref{tpz_Gamma_q}, since the mapping between the two is not a local single-photon basis change: the kernel $\Gamma$ couples the two photons collectively, leading to the perfectly correlated form in Eq. \eqref{tpz_Gamma_q}.

For collinear joint two-photon detection, a Zernike-mode measurement basis onto which the propagated SPDC state will be projected is
\begin{align}
    |\phi\rangle &=
    \int\d^2q \widetilde{Z}_{n_1}^{m_1}(\bq)\widetilde{Z}_{n_2}^{m_2}(\bq) a^\dagger(\bq) a^\dagger(\bq)|0\rangle
    \nn&=
    4\sum_{\substack{nn'n''\\mm'm''}} \Gamma_{n_1n_2n}^{m_1m_2m}\, 
    \Gamma_{n'n''n}^{m'm''m*}
    |\Tilde{z}_{n'}^{m'}, \Tilde{z}_{n''}^{m''}\rangle,
\end{align}
where we used Eqs. \eqref{eq:kintner_janssen} 
and \eqref{eq:Gamma_def}.
Note that the expansion coefficients of \eqref{tpz_Gamma} are given by $_{A}\langle\psi|\phi\rangle$:
\begin{align}
    _{A}\langle\psi|\phi\rangle&=
    8\sum_{\substack{nn'n''\\mm'm''}} \Gamma_{n_1n_2n}^{m_1m_2m}\, \Gamma_{n'n''n}^{m'm''m*}\, A_{n'n''N}^{m'm''M} \nn&= 
    8\,\Gamma_{n_1n_2N}^{m_1m_2M},\label{GGA_overlap}
\end{align}
following from the use of the identity in \eqref{GA_orth}, which now provides the link between non-collinear and collinear Zernike-mode structures.

To connect the propagated two-photon state with measurable quantities, we construct the two-photon detection amplitude
\begin{align}
    \mathcal{A}(\bro_1,\bro_2) &= 
    \langle 0\vert\hat \qE^{(+)}(\bro_2)\hat
    \qE^{(+)}(\bro_1)\vert\psi\rangle \label{tpaDef}
\end{align}
with the help of Eqs. \eqref{qE-Fhfr-turb} and \eqref{tpz}:
\begin{align}
    \mathcal{A}(\bro_1,\bro_2)\propto  \sum_{\substack{m_1n_1\\m_2n_2}}\!\!\zeta_{\,n_1n_2}^{m_1m_2}\,
    \mathcal J_{m_1n_1}(\bro_1)\,\mathcal J_{m_2n_2}(\bro_2), \label{tpa_expansion}
\end{align}
with the proportionality constant $-k^2e^{2ikz}/(2z^2\pi^3)$ and mode-filtered turbulent propagators
\begin{align}
    \mathcal J_{mn}(\br)& \equiv
    \int \d^2s \, 
    Z^m_n(\bs)\,
    e^{i\, \frac{k}{z} \bs \cdot\bro+\psi(\bs,\bro)}.
\end{align}
\eqref{tpa_expansion} is the far-field representation of the two-photon detection amplitude. 
In the thin-crystal approximation SPDC, for a pump profile prepared in a single Zernike mode $Z_N^M$, the detection amplitude simplifies considerably:
\begin{align}
    \mathcal{A}_{N}^{M}&(\bro_1,\bro_2)\propto \sum_{\substack{m_1n_1\\m_2n_2}}\!\!
    A_{\,n_1n_2N}^{m_1m_2M}
    \mathcal J_{m_1n_1}(\bro_1)\,\mathcal J_{m_2n_2}(\bro_2)
    \nn&=
    \int \d^2s \,Z^M_N(\bs)\,
    e^{i \frac{k}{z}\bs\cdot(\bro_1+\bro_2)+ \psi(\bs,\bro_1)+ \psi(\bs,\bro_2)},
    \label{tpa_expansion_Zpump}
\end{align}
where we used \eqref{A_Z_proj}.

\subsection{No-turbulence limit and recovery of exact selection rules.}
\st{Before analyzing the effects of atmospheric turbulence, it is instructive to consider the no-turbulence limit, in which the stochastic phase perturbation vanishes, $\psi(\bs,\bro)=0$. In this case,} Before evaluating the turbulence integrals, we first verify that our far-field detection amplitude correctly recovers the exact free-space Zernike selection rules previously established in Ref. \cite{Avetisyan25Z}. By taking the no-turbulence limit, where the stochastic phase perturbation vanishes ($\psi(\mathbf{s},\boldsymbol{\rho})=0$),
free-space propagation reduces to deterministic Fraunhofer diffraction, and the propagated field operators preserve the Zernike-mode operator expansion introduced in Section \ref{formalism}.
\begin{align}
    \mathcal{A}_{N}^{M}(\bro_1,\bro_2)&\propto 
    \sum_{\substack{m_1n_1\\m_2n_2}}\!\!
    A_{\,n_1n_2N}^{m_1m_2M}
    \widetilde Z_{n_1}^{m_1}\!\left(\bro_1\right)\,\widetilde Z_{n_2}^{m_2}\!\left(\bro_2\right) 
    \nn&= 
    \widetilde Z_N^M\!\left(\bro_1+\bro_2\right),
    \label{tpa_no_turb}
\end{align}
where the identity (\ref{F_A_Z_proj}) 
has been applied in the second equality. 
Eq. \eqref{tpa_no_turb} show that in the absence of turbulence the two-photon detection amplitude reduces to 
a correlation beam \cite{walborn} governed by the $A-$coefficients. 

The joint probability for a pair of photons in modes $z_{N_1}^{M_1}$ and $z_{N_2}^{M_2}$ can be calculated as follows \cite{Sonja}
\begin{align}
    &P_{N}^{M}\Big(z_{N_1}^{M_1}, z_{N_2}^{M_2}\Big) \propto
    \nn& 
    \left\langle\left |\iint \d^2\rho_1 \d^2\rho_2 \widetilde{Z}^{M_1\ast}_{N_1}(\bro_1)\widetilde{Z}^{M_2\ast}_{N_2}(\bro_2)
    \mathcal{A}_N^M(\bro_1,\bro_2) \right |^2\right\rangle.\label{eq:P-def}
\end{align}
The normalization factor is
\begin{align}
    \mathcal{N} = 
    \frac{1}{\sqrt{| \langle z_{n_1}^{m_1}|\psi\rangle|^2 \, | \langle z_{n_2}^{m_2}|\psi\rangle|^2}}.
\end{align}
Without turbulence, \eqref{eq:P-def} reduces, as it should, to (see appendix \ref{ap:no-turb})
\begin{align}
    P^\text{no turb}_{NM}\Big(z_{N_1}^{M_1},\,  z_{N_2}^{M_2}\Big) &= 
    \begin{cases}
    \left|A_{N_1N_2N}^{M_1M_2M}\right|^2, \quad \bro_1\neq\bro_2, \\\\
    \left|\Gamma_{N_1N_2N} ^{M_1M_2M}\right|^2, \quad \bro_1=\bro_2.
    \end{cases}
    \label{eq:P-final-noturb}
\end{align}
Recall from the representations given in Appendix \ref{app:A_defs},
\begin{align*}
    A_{N_1N_2N}^{M_1M_2M}&=0 \quad \text{for} \quad N>N_1+N_2, \text{whereas}\\
    \Gamma_{N_1N_2N}^{M_1M_2M}&=0 \quad \text{for} \quad N<N_1+N_2.
\end{align*} 
Accordingly, in the far-field, for a pump prepared in a single Zernike mode $Z_N^M$, the mode-coupling structure enforces exact selection rules: conservation of azimuthal order $M_1+M_2=M$, together with the radial constraint $N\leq N_1+N_2$ for the $A-$coefficients, and $N\geq N_1+N_2$ for the $\Gamma$ (subject to the usual parity condition). 

\subsection{Ensemble-averaged two-photon detection probability in turbulence.}
We now incorporate the effects of atmospheric turbulence by performing an ensemble average over the random phase perturbation $\psi(\bro,\bs)$. 
The following analysis assumes weak-to-moderate turbulence, such that the extended Huygens–Fresnel approximation remains valid and ensemble averages are dominated by second-order phase statistics; technical details are given in Appendix \ref{final_deriv}.
In contrast to the no-turbulence case, the joint two-photon detection probability is no longer determined solely by deterministic Fraunhofer propagation, but depends on statistical correlations of the turbulent medium.

The probability $P_{N}^{M}\Big(Z_{N_1}^{M_1}, Z_{N_2}^{M_2}\Big)$ for the case of collinear two photon amplitude $\mathcal{A}$ in thin crystal approximation is calculated in Appendix \ref{final_deriv}:
\begin{align}
    P_{N}^{M}\Big(z_{N_1}^{M_1}, z_{N_2}^{M_2}\Big) &=
    \sum_{n_1n_2} F_{n_1}F_{n_2}^*
    \sum_{n_5}G_{n_5}^0(\gamma,z) 
    A_{n_1n_2n_5}^{m_1,-m_10},\label{eq:P-final}
\end{align}
where 
\begin{align}
    F_n=\sum_{n'}
    \Gamma_{N_1N_2n'}^{M_1M_2,M_1+M_2}  \Gamma_{n'Nn}^{-M_1-M_2,M,M-M_1-M_2*},
\end{align}
and
\begin{align}
    G_{n_5}^{m_5}&(\gamma,z) = 
    \int \d^2u \,
    e^{-2\frac{\gamma k}{z}u^2}
    \widetilde{Z}_{n_5}^{m_5}\left(\frac{4kR}{z}\bu\right)
    \nn&=
    2\pi\delta_{m_50} \,  i^{n_5} \sqrt{n_5+1}\frac{\pi z}{4\gamma k}\left(\sqrt{\frac{8\pi^2kR^2}{\gamma z}}\right)^{n_5}
    \nn&\times
    \frac{\Gamma\left(\frac{n_5+2}{2}\right)}{\Gamma(n_5+2)}\,
    _1F_1\left(\frac{n_5+2}{2};n_5+2; -\frac{8\pi^2k R^2}{\gamma z}\right),
\end{align}
and $_1F_1$ is the confluent hypergeometric function, with $\gamma \equiv 0.4 \left(\sigma_R^2\right)^{6/5}$. Here, $\sigma_R^2 = 1.23 C_n^2 k^{7/6} z^{11/6}$ is the Rytov variance, which quantifies the accumulated turbulence strength over the propagation distance $z$ \cite{Andrews}. As detailed in Appendix \ref{final_deriv}, this variance emerges naturally from the ensemble average of the random phase perturbations $\psi(\bro, \bs)$ evaluated using the Kolmogorov spectrum.

\begin{figure}[t]
    \centering
    \begin{subfigure}[t]{0.9\linewidth}
        \centering
        \includegraphics[width=\linewidth]{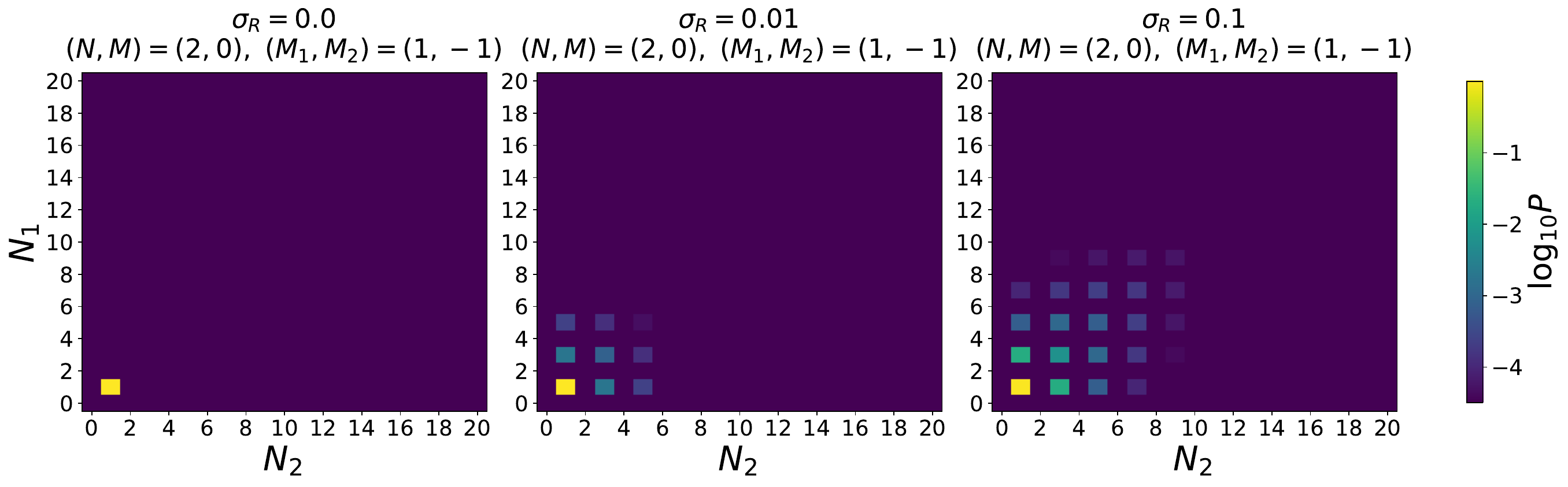}
        \label{fig:azimuthal_20}
    \end{subfigure}
    \hfill
    \begin{subfigure}[t]{0.9\linewidth}
        \centering
        \includegraphics[width=\linewidth]{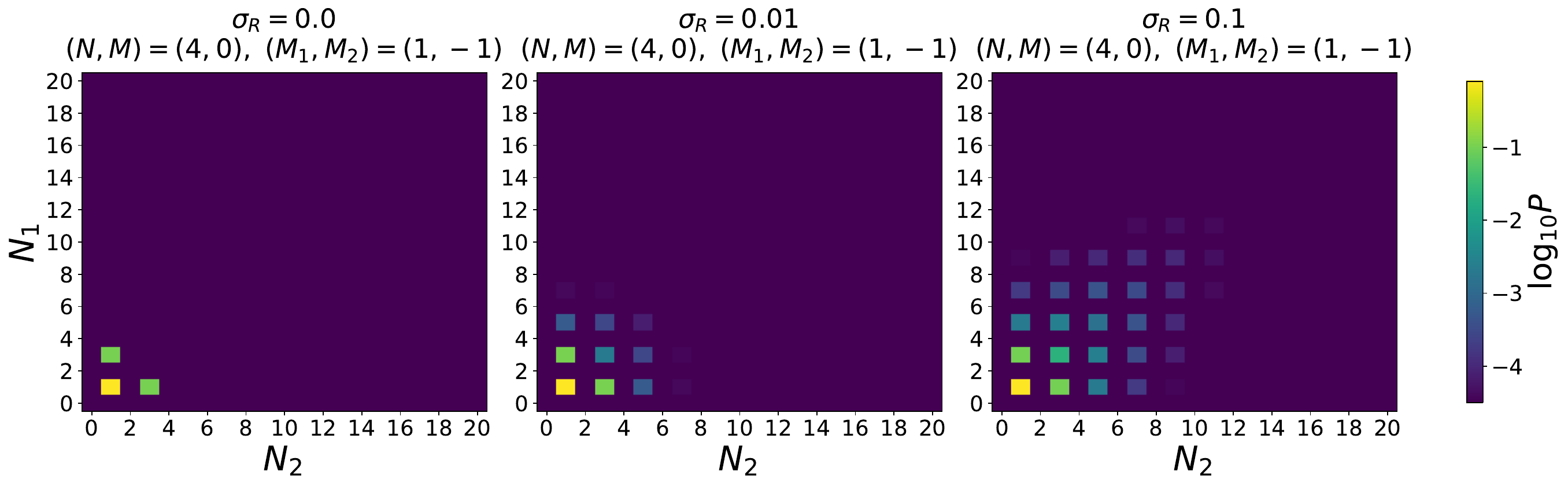}
        \label{fig:azimuthal_40}
    \end{subfigure}
    \caption{\justifying  
    Radial-mode correlations in Zernike-entangled photon pairs under varying turbulence strengths for fixed detector azimuthal indices $(M_1,M_2)=(1,-1)$. Shown is the base-10 logarithm of the normalized joint detection probability. The color scale is lower-bounded at $10^{-4.5}$ to isolate physically meaningful crosstalk. 
    \textbf{Top Row:} Pump mode $(N,M) = (2,0)$. \textbf{Bottom Row:} Pump mode $(N,M) = (4,0)$. 
    \textbf{Left} ($\sigma_R = 0.0$): In a perfect vacuum, exact momentum-matching and Zernike selection rules perfectly restrict the transition to a strictly bounded set of allowed states dictated by the free-space spatial overlap tensor. 
    \textbf{Center and Right} ($\sigma_R = 0.01$ and $0.1$): Increasing turbulence relaxes the radial-order constraints reflecting the  hierarchically bounded relaxation of the strict free-space Zernike constraints. The resulting crosstalk remains highly structured and securely confined to neighboring radial modes across both pump profiles. This demonstrates that bounded radial diffusion is a general feature of the Zernike basis rather than an artifact of a specific state preparation.
    \textit{Parameters}: $k=10^6 \text{ m}^{-1}$, $z=5\times 10^3 \text{ m}$, $R=5\times 10^{-3} \text{ m}$.
}
    \label{fig:heatmap}
\end{figure}

\section{Discussion}\label{discuss}
\st{A central result of this work is the exact analytical reduction of the continuous extended Huygens-Fresnel propagation integrals into a discrete, finite-dimensional algebraic framework. This formulation provides deep physical insight into a closed-form discrete algebraic framework.}
A central result of this work is the exact analytical reduction of the continuous extended Huygens-Fresnel propagation integrals into a discrete algebraic framework. While the derived algebraic framework is exact, numerical evaluation of equations such as Eq. (46) requires truncating the infinite sums over Zernike indices. For our numerical calculations, a finite radial cutoff is imposed. Convergence is ensured by incrementally increasing the maximum radial order until the normalized joint detection probabilities stabilize to a predetermined precision.

In the ideal vacuum limit ($\sigma_R \to 0$), the turbulence tensor $G_{n_5}^0(\gamma,z)$ reduces analytically to the origin-evaluation of the spatial Zernike polynomials. Within our algebraic framework, this rigorously collapses the probability down to a single $|\Gamma_{N_1 N_2 N}^{M_1 M_2 M}|^2$. As shown in the $\sigma_R = 0.0$ panel of Figure \ref{fig:heatmap}, this perfectly recovers the exact momentum-matching constraints of free-space diffraction: strict azimuthal conservation ($M = M_1 + M_2$) and the radial triangle inequality ($N \ge N_1 + N_2$). The absence of numerical artifacts in this limit highlights the exactness of the discrete tensor collapse.

When atmospheric turbulence is introduced ($\sigma_R > 0$), the discrete structure clearly delineates the breakdown of global selection rules. Because a single realization of a turbulent phase screen physically breaks the azimuthal symmetry of the propagating wavefront, the macroscopic azimuthal selection rule is violated ($M \neq M_1 + M_2$). This breakdown represents the onset of orbital angular momentum (OAM) crosstalk, a well-documented source of decoherence in turbulent free-space channels \cite{PhysRevLett.94.153901, Tyler:09}. However, while prior studies utilizing LG modes observe this OAM crosstalk as a broad, diffusive spread across azimuthal indices \cite{Tyler:09}, our formalism reveals that in the Zernike basis, this scattering remains hierarchically structured. The azimuthal leakage is explicitly routed by the $\Gamma$ tensors and geometrically bounded by the spatial overlap $A-$tensors. This sequence of tensor contractions explicitly governs the breakdown of macroscopic OAM conservation, quantifying the resulting turbulence-induced azimuthal crosstalk.

As visualized in Figure \ref{fig:heatmap} (Top Row), the probability is redistributed along increasing total radial orders, showing an outward diagonal decay along $N_1 + N_2$. Unlike the diffusive, unbounded mode-mixing observed in Hermite- or Laguerre-Gaussian bases, where atmospheric phase distortions scatter OAM indiscriminately across a broad range of azimuthal indices \cite{PhysRevLett.94.153901, Tyler:09}, the Zernike mode coupling is hierarchically constrained. Our algebraic framework mathematically guarantees that this crosstalk is dictated by the $\Gamma$-tensors and strictly constrained by the selection rules of the spatial overlap $A$-tensors.

To demonstrate that this structured degradation is a fundamental feature of the Zernike basis rather than an artifact of a specific state preparation, we analyze the transition probabilities for varying initial pump profiles. As illustrated in Figure \ref{fig:heatmap}—which evaluates a higher-order pump mode (e.g., $N=4, M=0$)—the physical behavior remains mathematically consistent across different states. The deterministic free-space constraints gradually break down into bounded radial crosstalk as the Rytov variance $\sigma_R^2$ increases. Because atmospheric turbulence is natively well-described by low-order aberrations, its dominant effects map highly efficiently onto the lowest-order $G$-tensor components. This fundamentally confines the leakage probability to adjacent macroscopic radial modes, preserving a highly structured correlation hierarchy that contrasts sharply with the uniform diffusion observed in standard OAM representations.
\st{ Because atmospheric turbulence is natively well-described by low-order aberrations (piston, tilt, astigmatism), its dominant effects map highly efficiently onto the lowest-order $G$-tensor components. This structured coupling implies that partial adaptive optics—targeting only the first few Zernike modes—can recover a disproportionately large fraction of the initial quantum spatial correlations.
Finally, a distinct advantage of our discrete algebraic framework is its native capacity to model partial adaptive optics (AO) without requiring computationally expensive phase-screen subtractions. In our formalism, the turbulence tensor $G_{n_5}^0(\gamma)$ acts as a discrete modal filter, dictating the statistical weight of the $n_5$-th aberration in driving intermodal crosstalk. To mathematically model an ideal AO system that perfectly compensates for low-order wavefront distortions up to a radial order $N_{AO}$, we simply apply a high-pass truncation to the transition probability network, restricting the sum in Eq. (46) to $n_5 > N_{AO}$. Figure 1 (Bottom Row) illustrates the physical consequence of this truncation. It shows the unaberrated baseline alongside the characteristic diagonal radial crosstalk induced by uncorrected turbulence. We apply a simulated AO correction up to primary spherical aberration ($N_{AO} = 6$). By merely filtering out these macroscopic, low-order modes, the extended diagonal crosstalk is almost entirely eliminated. Even when subjected to stronger turbulence ($\sigma_R = 0.5$), the correlation peak at $(N_1=1, N_2=1)$ remains intact. Notably, the residual high-frequency turbulence only drives faint scatterings into adjacent modes, proving that the bulk of two-photon decoherence in the Zernike basis is strictly driven by the lowest spatial frequencies of the Kolmogorov spectrum.}

\section{Conclusion}\label{summary}
We have developed an exact, discrete algebraic framework for the propagation and detection of Zernike-entangled two-photon states through atmospheric turbulence. By analytically resolving the extended Huygens-Fresnel integrals into a discrete sequence of spatial ($A$) and Fourier-domain ($\Gamma$) overlaps, and turbulence ($G$) tensors, we bypassed the severe numerical limitations of highly oscillatory continuous integrals. Our results demonstrate that turbulence-induced decoherence in the Zernike basis is highly structured. We explicitly showed how the macroscopic breakdown of OAM conservation mathematically coexists with the statistical isotropy of the turbulent medium. Furthermore, we showed that the degradation hierarchy is heavily dominated by low-order modes, reflecting the aberration-like nature of atmospheric phase distortions. These findings establish the Zernike basis not merely as a mathematical alternative to LG/HG modes, but as a naturally optimal representation for characterizing and mitigating decoherence in free-space quantum communication channels. Future work will extend this algebraic framework to strong-turbulence regimes and broadband sources  as well as model partial adaptive optics, and formally characterize the turbulent atmosphere as a high-dimensional quantum Zernike channel.
\st{Furthermore, we demonstrated that this discrete tensor formalism provides a closed algebraic method for modeling partial adaptive optics. By applying a simple high-pass truncation to the discrete turbulence tensor ($G_{n_5}^0$), we showed that correcting aberrations up to order $N_{AO}=6$ is sufficient to almost entirely suppress widespread radial crosstalk, even in stronger turbulence regimes ($\sigma_R = 0.5$). This confirms that quantum decoherence in the Zernike basis is heavily dominated by macroscopic, low-order phase distortions, offering a highly efficient theoretical blueprint for optimizing free-space quantum channels.}

\textbf{Acknowledgment.} We thank Varazdat Stepanyan for discussions.  
This work was supported by HESC (Higher Education and Science Committee) of Armenia, Grants No. 24FP-1F030 and 23/2IRF-1C003.

\bibliography{references}

\appendix
\onecolumngrid

\section{Derivation and summary of Zernike and Fourier–Zernike identities} \label{ap:identities}

\subsection{Representations of $A$ and $\Gamma$}
\label{app:A_defs}
For practical evaluation, both $A-$ and $\Gamma-$sets admit compact representations in terms of Clebsch--Gordan coefficients ($A$) and triple-Bessel integrals ($\Gamma$), which we summarize next.

\noindent\textit{Representation for $A$.}
The coefficients $A$ are given by the Clebsch-Gordan coefficients \cite{varshal, tango}
\begin{align}
    A_{n_1n_2n_3}^{m_1m_2m_3}= \sqrt{\frac{(n_1+1)(n_2+1)}{n_3+1}}\left|C_{\frac{n_1}{2}\frac{m_1}{2}\frac{n_2}{2}\frac{m_2}{2}}^{\frac{n_3}{2}\frac{m_3}{2}}\right|^2,
\end{align}
which are non-vanishing only when $n_1, n_2, n_3$ are non-negative integers or half-integers,
such that  $n_1+n_2+n_3$ is even
while satisfying the triangle conditions $|n_r - n_s| \leq n_t \leq n_r + n_s$
for any permutation $r,s, t$ of $1, 2, 3,$ and when $m_r = -n_r, -n_r + 1, \ldots, n_r - 1, n_r, \quad r = 1, 2, 3,$
with $m_1 + m_2 - m_3 = 0.$

\noindent\textit{Representation for $\Gamma$.}
The $\Gamma-$coefficients are evaluated as \cite{Janssen2011}: 
\begin{align}
    \Gamma_{N_1N_2n}^{mm'm''} &= \frac{\delta_{m'',m+m'} \, i^{N_1+N_2-n}}{\pi} \sqrt{\frac{n+1}{(N_1+1)(N_2+1)}}
    \nn&\times
    \left[Q_{N_1N_2}^{n+1}(1,1,2)+ Q_{N_1+2,N_2}^{n+1}(1,1,2)+ Q_{N_1,N_2+2}^{n+1}(1,1,2)+ Q_{N_1+2,N_2+2}^{n+1}(1,1,2)\right],\label{Gammas}
\end{align}
with $N_1+N_2-n$ even, which implies that $\Gamma$s are also real, and $Q_{ij}^k(a,b,c) = \int_0^\infty \d u J_i(au) J_j(bu) J_k(cu),$
so that
\begin{align}
    Q_{N_1N_2}^{n+1}(1,1,2) = 
    \begin{cases}
        \frac{\left(\tfrac{1}{2}(n+N_1+N_2)\right)!\left(\tfrac{1}{2}(n-N_1-N_2)\right)!}{\left(\tfrac{1}{2}(n-N_1+N_2)\right)!\left(\tfrac{1}{2}(n+N_1-N_2)\right)!} \,\, \frac{1}{2^{N_1+N_2+1}} \,\, P^{(N_1,N_2)}_{\frac{n-N_1-N_2}{2}}\!(0) \,\, P^{(N_2,N_1)}_{\frac{n-N_1-N_2}{2}}\!(0), \quad n \geq N_1+N_2, \\
        0, \quad n < N_1+N_2,
    \end{cases}\label{3Bessel}
\end{align}
\(P_k^{(\alpha,\beta)}\) being the Jacobi polynomial.

\subsection{New identities}

Using the completeness of the Zernike polynomials and the definition \eqref{A_def} of the $A-$coefficients,
the following summed product collapses to a $Z_N^M$ projector:
\begin{align}
&\sum_{\substack{n_1,m_1\\n_2,m_2}}
A_{n_1 n_2 N}^{m_1 m_2 M}\;
Z_{n_1}^{m_1}(\mathbf s_1)\,Z_{n_2}^{m_2}(\mathbf s_2)
= \frac{1}{\pi}\int \d^2 s\, Z_N^M(\mathbf s)
  \Big[\sum_{n_1,m_1} Z_{n_1}^{m_1}(\mathbf s_1)\,Z_{n_1}^{m_1}(\mathbf s)^{*}\Big]
  \Big[\sum_{n_2,m_2} Z_{n_2}^{m_2}(\mathbf s_2)\,Z_{n_2}^{m_2}(\mathbf s)^{*}\Big]
  \nn&= 
  \int \d^2 s\, Z_N^M(\mathbf s)
  \Big[\sum_{n_1,m_1} Z_{n_1}^{m_1}(\mathbf s_1)\,Z_{n_1}^{m_1}(\mathbf s)^{*}\Big]
  \delta_{D}(\mathbf s_2-\mathbf s)=
  \pi\,
    Z_N^M(\mathbf s_2)\,\Big[\sum_{n_1,m_1} Z_{n_1}^{m_1}(\mathbf s_1)\,Z_{n_1}^{m_1}(\mathbf s_2)^{*}\Big]\,
    = 
    \pi\,
    Z_N^M(\mathbf s_1)\,\delta_{D}(\mathbf s_1-\mathbf s_2).\label{a14}
\end{align}
The identity should be understood in the distributional sense under integration against admissible disk-supported test functions.
In the Fourier domain, we have
\begin{align}
    \sum_{\substack{n_1,m_1\\n_2,m_2}}&
    A_{n_1 n_2 N}^{m_1 m_2 M}\;
    \widetilde{Z}_{n_1}^{m_1}(\bro_1)\,
    \widetilde{Z}_{n_2}^{m_2}(\bro_2)
    = \pi \widetilde{Z}_{N}^M(\bro_1+\bro_2).
\end{align}
Similarly, using the definition \eqref{eq:Gamma_def} of $\Gamma$ 
we have
\begin{align}
     \sum_{\substack{n_1,m_1\\n_2,m_2}} \Gamma_{n_1n_2N}^{m_1m_2M}  \widetilde{Z}_{n_1}^{m_1*}(\bq_1) \widetilde{Z}_{n_2}^{m_2*}(\bq_2) &= \frac{1}{\pi}\int \d^2q \, 
     \widetilde{Z}_{N}^{M*}(2\bq)
     \left[\sum_{n_1m_1}
     \widetilde{Z}_{n_1}^{m_1}(\bq)\widetilde{Z}_{n_1}^{m_1*}(\bq_1) \right]
     \left[\sum_{n_2m_2}\widetilde{Z}_{n_2}^{m_2}(\bq)\widetilde{Z}_{n_2}^{m_2*}(\bq_2)\right]
     \nn&=
     \pi\int \d^2q \,
     \widetilde{Z}_{N}^{M*}(2\bq)\,\widetilde{Z}_0^0(\bq-\bq_1)\,\widetilde{Z}_0^0(\bq-\bq_2)
     .\label{ap:F_G_Z_proj}
\end{align}
As the Fourier transform of the product $ \widetilde{Z}_{N}^{M*}(2\bq)\,\widetilde{Z}_0^0(\bq-\bq_1)$ is not supported on the unit disk, the projector $\widetilde{Z}_0^0(\bq-\bq_2)$ does not act as a delta function. 
Now, double Fourier transforming \eqref{ap:F_G_Z_proj}, we get
\begin{align}
     \sum_{\substack{n_1,m_1\\n_2,m_2}} \Gamma_{n_1n_2N}^{m_1m_2M} Z_{n_1}^{m_1*}(\bs_1) Z_{n_2}^{m_2*}(\bs_2) &=
     \pi\int \d^2q\, e^{-2\pi i \bq\cdot(\bs_1+\bs_2)}\widetilde{Z}_{N}^{M*}(2\bq)=
     \frac{\pi}{4}Z_N^{M*}\!\left(\frac{\bs_1+\bs_2}{2}\right), \quad |\bs_1|\leq 1, ~~ |\bs_2|\leq 1.
\end{align}

\section{No turbulence limit.}\label{ap:no-turb}
We calculate the joint probability \eqref{eq:P-def} using \eqref{tpa_no_turb} in thin crystal approximation and collinear cases.
\paragraph{Thin crystal approximation.}
\begin{align}
    P_{N}^{M}\Big(z_{N_1}^{M_1}, z_{N_2}^{M_2}\Big) 
    &\propto
    \left |\int \d^2\rho_1 \int\d^2\rho_2 \, \widetilde{Z}^{M_1\ast}_{N_1}(\bro_1)\widetilde{Z}^{M_2\ast}_{N_2}(\bro_2)
    \mathcal{A}_N^M(\bro_1,\bro_2) \right |^2
    \nn&=
    \Big |    
    \int \d^2\rho_1 
    \widetilde{Z}^{M_1\ast}_{N_1}(\bro_1)
    \int\,\d^2\rho_2 
    \widetilde{Z}_N^M(\bro_1+\bro_2) 
    \widetilde{Z}^{M_2\ast}_{N_2}(\bro_2)
    \Big |^2
    \nn&=
    \Big |    
    \int \d^2\rho \, 
    \widetilde{Z}_N^M(\bro)
    \int\,\d^2\rho_1 
    \widetilde{Z}^{M_1\ast}_{N_1}(\bro_1) 
    \widetilde{Z}^{M_2\ast}_{N_2}(\bro-\bro_1)
    \Big |^2, \qquad (\bro=\bro_1+\bro_2,\, \bro_1=\bro_1)
    \nn&=
    \Big |    
    \int \d^2\rho  
    \widetilde{Z}^{M}_{N}(\bro) 
    \left(\widetilde{Z}_{N_1}^{M_1\ast}\ast
    \widetilde{Z}^{M_2\ast}_{N_2}\right)(\bro)
    \Big |^2 =
    \Big | \sum_{N_3}A_{N_1N_2N_3}^{M_1,M_2,M_1+M_2}    
    \int \d^2\rho  
    \widetilde{Z}^{M}_{N}(\bro) 
    \widetilde{Z}^{M_1+M_2\ast}_{N_3}(\bro)
    \Big |^2
    \nn&=
    \Big |    
    \sum_{N_3}A_{N_1N_2N_3}^{M_1M_2,M_1+M_2}
    \delta_{NN_3}\delta_{M,M_1+M_2}
    \Big |^2
    =\left|A_{N_1N_2N}^{M_1M_2M}\right|^2,
\end{align}
where we used \eqref{tpa_no_turb} in the first line, and the Fourier transform of \eqref{A_cg} in the fourth line. 
\paragraph{$\bro_1=\bro_2=\bro$ case.}
\begin{align}
    P_{N}^{M}\Big(z_{N_1}^{M_1}, z_{N_2}^{M_2}\Big) 
    &\propto
    \left |\int \d^2\rho \widetilde{Z}^{M_1\ast}_{N_1}(\bro)\widetilde{Z}^{M_2\ast}_{N_2}(\bro)
    \mathcal{A}_N^M(\bro,\bro) \right |^2=
    \Big |    
    \int \d^2\rho 
    \widetilde{Z}^{M_1\ast}_{N_1}(\bro)
    \widetilde{Z}^{M_2\ast}_{N_2}(\bro)
    \widetilde{Z}_N^M(2\bro) 
    \Big |^2 \propto 
    \left|\Gamma_{N_1N_2N}^{M_1M_2M}\right|^2.
\end{align}
The above results are consistent with Eq. (\ref{tpz}) for a single $Z_N^M$ pump profile.

\section{Derivation of \eqref{eq:P-final}} \label{final_deriv}
\paragraph{Continuous Spatial Representation:}
We write the joint probability \eqref{eq:P-def} 
and consider, for simplicity, the case $\bro_1=\bro_2=\bro$: 
\begin{align}
    P_{N}^{M}\Big(z_{N_1}^{M_1}, z_{N_2}^{M_2}\Big) &= 
    \frac{k^4}{z^4}
    \Bigg\langle\Big |
    \int \d^2\rho \widetilde{Z}^{M_1\ast}_{N_1}\left(\frac{kR}{z}\bro\right)\widetilde{Z}^{M_2\ast}_{N_2}\left(\frac{kR}{z}\bro\right) \, 
    \int \d^2s \,
    Z_{N}^{M}\left(\frac{\bs}{R}\right)\,
    e^{2\pi i\, \tfrac{2k}{z}\bs\cdot\bro} 
    e^{2\psi(\bs,\bro)}
    \Big |^2
    \Bigg\rangle 
    \nn& = 
    \frac{k^4}{z^4}
    \iint \d^2\rho_1 \d^2\rho_2 
    \widetilde{Z}_{N_1}^{M_1*}\left(\frac{kR}{z}\bro_1\right)\, \widetilde{Z}_{N_2}^{M_2*}\left(\frac{kR}{z}\bro_1\right)\, 
    \widetilde{Z}_{N_1}^{M_1}\left(\frac{kR}{z}\bro_2\right)\, \widetilde{Z}_{N_2}^{M_2}\left(\frac{kR}{z}\bro_2\right)
    \nn&\qquad\times 
    \iint \d^2s_1\d^2s_2\,
    Z_{N}^{M}\left(\frac{\bs_1}{R}\right)\,
    Z_{N}^{M*}\left(\frac{\bs_2}{R}\right)\,
    e^{2\pi i\,\frac{2k}{z} (\bs_1 \cdot\bro_1-\bs_2\cdot\bro_2)}
    \left\langle 
    e^{2[\psi(\bs_1,\bro_1)+ 
    \psi^{\ast}(\bs_2, \bro_2)]}
    \right\rangle,
\end{align}
where we also write the various quantities in dimensionless form ($\bro\to\frac{kR}{z}\bro$, $\bs\to\frac{\bs}{R}$).
Using the method of cumulants up to second order,
$\langle\exp(\psi)\rangle = \exp\left(\langle\psi\rangle+\frac{1}{2}\left(\langle\psi^2 \rangle-\langle\psi\rangle^2\right)\right),$
we write, by definition, $\left\langle \exp\left[2( \psi(\bs_1,\bro_1)+ 
\psi^\ast(\bs_2, \bro_2) )\right]\right\rangle=
\exp\left[- 2 D(\bro_1-\bro_2,\bs_1-\bs_2)\right],$
where $D(\bp,\bQ) = 8\pi k^2z\int_0^1\d \xi\int_0^\infty \d\kappa\,\kappa \, \Phi(\kappa)\left(1-J_0[|(1-\xi )\bp+\xi\bQ|\kappa\right)$ 
which evaluates to $D(\bp,\bQ) =\frac{\gamma k}{z}(p^2+\bp\cdot\bQ+Q^2),
$ \cite[Ch. 7]{Andrews}, for Kolmogorov spectrum $\Phi(\kappa)$,
with $\gamma \equiv 0.4 \left(\sigma_R^2\right)^{6/5}.$
We now use the identity \eqref{eq:F_kintner_janssen} and simplify the expression of the probability further:
\begin{align}
    P_{N}^{M}&\Big(z_{N_1}^{M_1}, z_{N_2}^{M_2}\Big) =
    \frac{k^4}{z^4}
    4\sum_{nm}\Gamma_{N_1N_2n}^{M_1M_2m*}
    4\sum_{n'm'}\Gamma_{N_1N_2n'}^{M_1M_2m'}
    \iint \d^2s_1\d^2s_2\,
    Z_{N}^{M}(\bs_1/R)\,
    Z_{N}^{M*}(\bs_2/R)\,
    \nn&\times 
    \iint \d^2\rho_1 \,\d^2\rho_2\,
    \widetilde{Z}_{n}^{m*}\left(2\frac{kR}{z}\bro_1\right)\,   
    \widetilde{Z}_{n'}^{m'}\left(2\frac{kR}{z}\bro_2\right)
    e^{2\pi i\,\frac{2k}{z}(\bs_1\cdot\bro_1- \bs_2\cdot\bro_2)}
    e^{-2\frac{\gamma k}{z}[|\bs_1-\bs_2|^2+|\bro_1-\bro_2|^2+(\bs_1-\bs_2)\cdot(\bro_1-\bro_2)]}.
\end{align}
Defining $\bQ = \bs_1-\bs_2$, $2\bS = \bs_1+\bs_2$, $\bp = \bro_1-\bro_2$ and $2\textbf{P} = \bro_1+\bro_2$, such that $\d^2s_1\,\d^2s_2=\d^2S\,\d^2Q$ and $\d^2\rho_1\,\d^2\rho_2 = \d^2p\,\d^2P$, the probability can be cast into
\begin{align}
    P_{N}^{M}\Big(z_{N_1}^{M_1}, z_{N_2}^{M_2}\Big) &=
    \sum_{nn'}  \Gamma_{N_1N_2n }^{M_1M_2m*}\,
    \Gamma_{N_1N_2n'}^{M_1M_2m'}\,
    I_{nn'N}^{mm'M},\label{ap:prob}
\end{align}
where $m=m'=M_1+M_2$ due to selection rules imposed by $\Gamma$s and 
\begin{align}
    I_{nn'N}^{mm'M}&=\frac{16k^4}{z^4} \int \d^2p \,
    \int \d^2Q \, \left[\int \d^2 S \,
    Z_N^M\left(\tfrac{1}{R}(\bS+\bQ/2)\right)
    Z_N^{M*}\left(\tfrac{1}{R}(\bS-\bQ/2)\right)
    e^{2\pi i\,\frac{2k}{z}\bS\cdot\bp}\right]
    \nn&\times
    e^{-2\frac{\gamma k}{z}[Q^2+p^2+\bQ\cdot\bp]}
    \left[\int d^2P \,
    \widetilde{Z}_{n}^{m*}\left(\tfrac{kR}{z}(2\textbf{P}+\bp)\right)\,   
    \widetilde{Z}_{n'}^{m'}\left(\tfrac{kR}{z}(2\textbf{P}-\bp)\right)
    e^{2\pi i\,\frac{2k}{z}\bQ\cdot \textbf{P}}\right].
\end{align}
\paragraph{Decoupling via Fourier-Zernike Addition:}
The integrals in square brackets (hereafter called $I_{N}^{M}(\bQ,\bp)$ and $I_{nn'}^{mm'}(\bQ,\bp)$, respectively) can be calculated analytically with the help of the addition theorem \eqref{F_A_Z_proj}. 
\begin{align}
    I_{nn'}^{mm'}(\bQ,\bp) = \int d^2 P \, 
    \widetilde{Z}_{n}^{m*}\left(\tfrac{kR}{z}(2\textbf{P}+\bp)\right)\,   
    \widetilde{Z}_{n'}^{m'}\left(\tfrac{kR}{z}(2\textbf{P}-\bp)\right)
    e^{2\pi i\,\frac{k}{z}2\bQ\cdot \textbf{P}}. 
\end{align}
Now define $\bv = \frac{kR}{z}\bp, \quad \bq=\tfrac{kR}{z}(2\textbf{P}-\bp),$ and use the identity \eqref{F_A_Z_proj}, then \eqref{eq:F_kintner_janssen}.
\begin{align}
    I_{nn'}^{mm'}(\bQ,\bp) &= 
    \frac{z^2}{4k^2R^2}\int d^2 q \, 
    \widetilde{Z}_{n}^{m*} (\bq+2\bv) \,   
    \widetilde{Z}_{n'}^{m'}(\bq) 
    e^{2\pi i\,\frac{1}{R}\bQ\cdot (\bq+\bv)} 
    \nn&=
    \frac{z^2}{4k^2R^2\pi}\sum_{\substack{n_1n_2\\m_1m_2}} A_{n_1n_2n}^{m_1m_2m*}\, \widetilde{Z}_{n_2}^{m_2*}(2\bv)\int d^2 q \,
    \widetilde{Z}_{n'}^{m'}\left(\bq\right)\,   
    \widetilde{Z}_{n_1}^{m_1*}\left(\bq\right)
    e^{2\pi i\,\frac{1}{R}\bQ\cdot (\bq+\bv)} 
    \nn&=
    \frac{z^2}{k^2R^2\pi}e^{2\pi i\,\frac{1}{R}\bQ\cdot\bv}  \sum_{\substack{n_1n_2\\m_1m_2}}A_{n_1n_2n}^{m_1m_2m*} \,\widetilde{Z}_{n_2}^{m_2*}(2\bv)
    \sum_{n_3m_3}\Gamma_{n'n_1n_3}^{m', -m_1 m_3} (-1)^{n_1} \int d^2 q \,
    \widetilde{Z}_{n_3}^{m_3}\left(2\bq\right)
    e^{2\pi i\,\frac{1}{R}\bQ\cdot\bq} 
    \nn&=
    \frac{z^2}{4k^2R^2\pi}e^{2\pi i\,\frac{k}{z}\bQ\cdot\bp}  \sum_{\substack{n_1n_2\\m_1m_2}} A_{n_1n_2n}^{m_1m_2m*} \, \widetilde{Z}_{n_2}^{m_2*}\left(\frac{2kR}{z}\bp\right)
    \sum_{n_3m_3}\Gamma_{n'n_1n_3}^{m',-m_1m_3}
    (-1)^{m_3+n_1}Z_{n_3}^{m_3}\left(\frac{\bQ}{2R}\right).
\end{align}
Similarly, we have 
\begin{align}
    I_{N}^{M}(\bQ,\bp)&=\int \d^2 S \,
    Z_N^M\left(\tfrac{1}{R}(\bS+\bQ/2)\right)
    Z_N^{M*}\left(\tfrac{1}{R}(\bS-\bQ/2)\right)
    e^{2\pi i\,\frac{k}{z}2\bS\cdot\bp}
    \nn&=
    \frac{R^2}{\pi}e^{-2\pi i\,\frac{k}{z}\bQ\cdot\bp}
    \sum_{\substack{n_1'n_2'n_3'\\m_1'm_2'm_3'}} (-1)^{n_1'+m_3'} A_{n_1'n_2'N}^{m_1'm_2'M} \, \Gamma_{Nn_1'n_3'}^{M,-m_1'm_3'*}
    \widetilde{Z}_{n_2'}^{m_2'}\!\left(\frac{2kR}{z}\bp\right)
    Z_{n_3'}^{m_3'*}\left(\frac{\bQ}{2R}\right).
\end{align}
Finally,
\begin{align}
I_{nn'N}^{mm'M}&=\frac{16k^4}{z^4}\int \d^2p \,
    \int \d^2Q \, 
    e^{-2\frac{\gamma k}{z}[Q^2+p^2+\bQ\cdot\bp]}
    I_{nn'}^{M_1M_2}(\bQ,\bp) I_{N}^{M}(\bQ,\bp) 
    \nn&=
    \frac{4k^2}{\pi^2z^2}
    \sum_{\substack{n_1n_2n_3\\ n_1'n_2'n_3'\\m_1m_2m_3\\m_1'm_2'm_3'}}  (-1)^{n_1'+m_3'+n_1+m_3} A_{n_1'n_2'N}^{m_1'm_2'M} \Gamma_{Nn_1'n_3'}^{M,-m_1'm_3'*} 
    A_{n_1n_2n}^{m_1m_2m*}\, \Gamma_{n'n_1n_3}^{m',-m_1m_3} 
    \nn&\times 
    \int \d^2p \,
    \int \d^2Q \, 
    e^{-2\frac{\gamma k}{z}[Q^2+p^2+\bQ\cdot\bp]}
    \widetilde{Z}_{n_2}^{m_2*}\left(\frac{2kR}{z}\bp\right)
    \widetilde{Z}_{n_2'}^{m_2'} \left(\frac{2kR}{z}\bp\right)
    Z_{n_3}^{m_3}\left(\frac{\bQ}{2R}\right)
    Z_{n_3'}^{m_3'*}\left(\frac{\bQ}{2R}\right).
\end{align}
Using again the definitions of $\Gamma$ and $A$ to linearize the Zernike products, we get
\begin{align}
I_{nn'N}^{mm'M}&=
    \frac{4k^2}{\pi^2z^2}
    \sum_{\substack{n_1n_2n_3\\ n_1'n_2'n_3'\\m_1m_2m_3\\m_1'm_2'm_3'}} (-1)^{n_1'+m_3'+n_1+m_3} A_{n_1'n_2'N}^{m_1'm_2'M} \Gamma_{Nn_1'n_3'}^{M,-m_1'm_3'*} 
    A_{n_1n_2n}^{m_1m_2m*}\, \Gamma_{n'n_1n_3}^{m',-m_1m_3}  
    \nn&\times
    4\sum_{\substack{n_4n_4'\\m_4m_4'}}(-1)^{n_2} \Gamma_{n_2n_2'n_4}^{-m_2m_2'm_4}\, A_{n_3n_3'n_4'}^{m_3,-m_3'm_4'}
    \int \d^2p \,
    \int \d^2Q \, 
    e^{-2\frac{\gamma k}{z}[Q^2+p^2+\bQ\cdot\bp]}
    \widetilde{Z}_{n_4}^{m_4}\left(\frac{4kR}{z}\bp\right)
    Z_{n_4'}^{m_4'}\left(\frac{\bQ}{2R}\right).\label{c11}
\end{align}
To evaluate the last integrals, defined as $\mathcal{T}_{n_4n_4'}^{m_4m_4'}$, change the variables $\bu = \bp+\frac{\bQ}{2}$ and use again Eq. \eqref{F_A_Z_proj},
\begin{align}
    \mathcal{T}_{n_4n_4'}^{m_4m_4'}&\equiv\int \d^2Q \, 
    e^{-\frac{3\gamma k}{2z}Q^2}
    Z_{n_4'}^{m_4'}\left(\frac{\bQ}{2R}\right)
    \int \d^2u \,
    e^{-2\frac{\gamma k}{z}u^2}
    \widetilde{Z}_{n_4}^{m_4}\left(\frac{4kR}{z}(\bu-\bQ/2)\right)
    \nn&=
    \sum_{\substack{n_5n_5'\\m_5m_5'}} A_{n_5n_5'n_4}^{m_5m_5'm_4}\,
    \int \d^2Q \, 
    e^{-\frac{3\gamma k}{2z}Q^2}
    Z_{n_4'}^{m_4'}\left(\frac{\bQ}{2R}\right)
    \widetilde{Z}_{n_5}^{m_5}\left(-\frac{2kR}{z}\bQ\right)\,
    \int \d^2u \,
    e^{-2\frac{\gamma k}{z}u^2}
    \widetilde{Z}_{n_5'}^{m_5'}\left(\frac{4kR}{z}\bu\right)
    .\label{c12}
\end{align} 
\paragraph{Evaluation of the Turbulence Tensor:} Define
\begin{align}
    G_{n_5'}^{m_5'}(\gamma) &= 
    \int \d^2u \,
    e^{-2\frac{\gamma k}{z}u^2}
    \widetilde{Z}_{n_5'}^{m_5'} \left(\frac{4kR}{z}\bu\right)=
    2\pi\delta_{m_5'0}\,i^{n_5'}\sqrt{n_5'+1} \,\frac{z}{4kR}\int_0^\infty \d u \,
    e^{-2\frac{\gamma k}{z}u^2}
    J_{n_5'+1}\left(\frac{8\pi kR}{z} u\right)
    \nn&=
    2\pi\delta_{m_5'0} \,  i^{n_5'} \sqrt{n_5'+1}\frac{\pi z}{4\gamma k}\left(\sqrt{\frac{8\pi^2kR^2}{\gamma z}}\right)^{n_5'}
    \frac{\Gamma\left(\frac{n_5'+2}{2}\right)}{\Gamma(n_5'+2)}\,\,
    _1F_1\left(\frac{n_5'+2}{2};n_5'+2; -\frac{8\pi^2k R^2}{\gamma z}\right)
    \label{c17}
\end{align}
where we used \cite[6.631-1]{Gradshteyn}) and the fact that, for even $n$, $Z_{n}^{0}(0)=\sqrt{n+1}(-1)^{n/2}=\sqrt{n+1}i^{n}$, and
\begin{align}
    \mathcal{G}_{n_4'n_5}^{m_4'm_5}(\gamma)&= \int \d^2Q \, 
    e^{-\frac{3\gamma k}{2z}Q^2}
    Z_{n_4'}^{m_4'}\left(\frac{\bQ}{2R}\right)
    \widetilde{Z}_{n_5}^{m_5}\left(-\frac{2kR}{z}\bQ\right).
    \label{c18}
\end{align}
As $\gamma \to 0$, $\,_1F_1(a; b; -x) \to \frac{\Gamma(b)}{\Gamma(b-a)} x^{-a}$, and $G_{n'_5}^{m_5'} \to \frac{z^2}{16k^2 R^2} Z_{n'_5}^{m_5'}(0)$.
The probability \eqref{ap:prob} takes the form
\begin{align}
    P_{N}^{M}\Big(z_{N_1}^{M_1}, z_{N_2}^{M_2}\Big) &=
    \left(\frac{4k}{\pi z}\right)^2
    \sum_{\substack{nn_1n_2n_3n_4n_5\\ n'n_1'n_2'n_3'n_4'n_5'\\m_1m_2m_3m_4m_5\\m_1'm_2'm_3'm_4'm_5'}} (-1)^{n_1'+m_3'+n_1+m_3+n_2}
    \Gamma_{N_1N_2n }^{M_1M_2m}\,
    A_{n_1n_2n}^{m_1m_2m*}\,
    \Gamma_{N_1N_2n'}^{M_1M_2m'*}\,
    \Gamma_{n'n_1n_3}^{m',-m_1m_3*}\,
    \nn&\qquad\qquad\qquad\times
    A_{n_3n_3'n_4'}^{m_3,-m_3'm_4'*}\,
    A_{n_1'n_2'N}^{m_1'm_2'M} \,  
    \Gamma_{n_2n_2'n_4}^{-m_2m_2'm_4*}\,
    \Gamma_{Nn_1'n_3'}^{M,-m_1'm_3'}\,
    A_{n_5n_5'n_4}^{m_5m_5'm_4}\,
    \mathcal{G}_{n_4'n_5}^{m_4',m_5} (\gamma) G_{n_5'}^{m_5'}(\gamma)
\end{align}
Substitute the continuous integral definitions of $A$ and $\Gamma$, and assigning spatial integration variables ($\bs, \bs_1, \bs_2, \bs_3$) to the four $A$-tensors, and frequency integration variables ($\bq, \bq_1, \bq_2, \bq_3, \bq_4$) to the five $\Gamma$-tensors. The intermediate summations naturally group pairs of spatial and Fourier Zernike polynomials sharing identical indices.
\begin{align}
    &P_{N}^{M}\Big(z_{N_1}^{M_1}, z_{N_2}^{M_2}\Big) =
    \left(\frac{4k}{\pi z}\right)^2 
    \frac{1}{\pi^9}
    \iiint \d^2q\, \d^2q_1 \, \d^2q_2
    \iint\d^2q_3\,\d^2q_4
    \iiint\d^2s \, \d^2s_1\, \d^2s_2\,\d^2s_3
    \nn&\times
    \widetilde{Z}_{N_1}^{M_1}(\bq)
    \widetilde{Z}_{N_2}^{M_2}(\bq)
    \widetilde{Z}_{N_1}^{M_1*}(\bq_1)
    \widetilde{Z}_{N_2}^{M_2*}(\bq_1)
    \widetilde{Z}_{N}^{M}(\bq_4)
    Z_{N}^{M*}(\bs_2)
    \nn&\times
    \underbrace{
    \left[
        \sum_{nm}
        Z_n^{m}(\bs)
        \widetilde{Z}_{n}^{m*}(2\bq)
    \right]}_{\pi e^{-2\pi i\bs\cdot 2\bq}}
    \underbrace{
    \left[
        \sum_{n'm'}
        \widetilde{Z}_{n'}^{m'}(2\bq_1)
        \widetilde{Z}_{n'}^{m'*}(\bq_2)
    \right]}_{\pi \widetilde{Z}_0^0(\bq_2-2\bq_1)}
    \underbrace{
    \left[
        \sum_{n_1m_1}(-1)^{n_1}
        Z_{n_1}^{m_1*}(\bs)
        \widetilde{Z}_{n_1}^{-m_1*}(\bq_2)
    \right]}_{\pi e^{2\pi i\bs\cdot \bq_2}}
    \nn&\times
    \underbrace{
    \left[
        \sum_{n_2m_2}(-1)^{n_2}
        Z_{n_2}^{m_2*}(\bs)
        \widetilde{Z}_{n_2}^{-m_2*}(\bq_3)
    \right]}_{\pi e^{2\pi i\bs\cdot \bq_3}}
    \underbrace{
    \left[
        \sum_{n_3m_3}(-1)^{m_3}
        Z_{n_3}^{m_3*}(\bs_1)
        \widetilde{Z}_{n_3}^{m_3}(2\bq_2)
    \right]}_{\pi e^{-2\pi i\bs_1\cdot 2\bq_2}}
    \underbrace{
    \left[
        \sum_{n_1'm_1'}(-1)^{n_1'}
        Z_{n_1'}^{m_1'}(\bs_2)
        \widetilde{Z}_{n_1'}^{-m_1'}(\bq_4)
    \right]}_{\pi e^{-2\pi i\bs_2\cdot \bq_4}}
    \nn&\times
    \underbrace{
    \left[
        \sum_{n_2'm_2'}
        Z_{n_2'}^{m_2'}(\bs_2)
        \widetilde{Z}_{n_2'}^{m_2'*}(\bq_3)
    \right]}_{\pi e^{-2\pi i\bs_2\cdot \bq_3}}
    \underbrace{
    \left[
        \sum_{n_3'm_3'}(-1)^{m_3'}
        Z_{n_3'}^{-m_3'*}(\bs_1)
        \widetilde{Z}_{n_3'}^{m_3'*}(2\bq_4)
    \right]}_{\pi e^{2\pi i\bs_1\cdot 2\bq_4}}
    \underbrace{
    \left[
        \sum_{n_4m_4}
        Z_{n_4}^{m_4*}(\bs_3)
        \widetilde{Z}_{n_4}^{m_4}(2\bq_3)\,
    \right]}_{\pi e^{2\pi i\bs_3\cdot 2\bq_3}}
    \nn&\times
    \sum_{\substack{n_4'n_5\\m_4'm_5}}
    \mathcal{G}_{n_4'n_5}^{m_4',m_5} (\gamma)
    Z_{n_4'}^{m_4'}(\bs_1)\,
    Z_{n_5}^{m_5}(\bs_3)
    \sum_{n_5'm_5'}
    G_{n_5'}^{m_5'}(\gamma)
    Z_{n_5'}^{m_5'}(\bs_3).
\end{align}
Substituting and rearranging, we get
\begin{align}
P_{N}^{M}\Big(& z_{N_1}^{M_1}, z_{N_2}^{M_2}\Big) =
    \left(\frac{4k}{\pi z}\right)^2 
    \iiint \d^2q\, \d^2q_1 \, \d^2q_3\, \d^2q_4
    \widetilde{Z}_{N_1}^{M_1}(\bq)
    \widetilde{Z}_{N_2}^{M_2}(\bq)
    \widetilde{Z}_{N_1}^{M_1*}(\bq_1)
    \widetilde{Z}_{N_2}^{M_2*}(\bq_1)
    \widetilde{Z}_{N}^{M}(\bq_4)
    \nn&\times
    \int \, \d^2s_2\,
    e^{-2\pi i \bs_2\cdot(\bq_4+\bq_3)}
    Z_{N}^{M*}(\bs_2)
    \underbrace{
    \int\d^2s \, 
    e^{-2\pi i \bs\cdot 2\bq}
    e^{2\pi i \bs\cdot 2\bq_1}
    e^{2\pi i \bs\cdot\bq_3}}_{\widetilde{Z}_0^0(2\bq-2\bq_1-\bq_3) }
    \sum_{\substack{n_4'n_5\\m_4'm_5}}
    \mathcal{G}_{n_4'n_5}^{m_4'm_5} (\gamma)
    \nn&\times
    \int \, \d^2s_1\,
    e^{2\pi i\bs_1\cdot (2\bq_4-4\bq_1)}
    Z_{n_4'}^{m_4'}(\bs_1)\,
    \int\d^2s_3\,
    e^{2\pi i \bs_3\cdot 2\bq_3}
    Z_{n_5}^{m_5}(\bs_3)
    \sum_{n_5'm_5'}
    G_{n_5'}^{m_5'}(\gamma)
    Z_{n_5'}^{m_5'}(\bs_3)
    \nn&=
    \left(\frac{4k}{\pi z}\right)^2 \frac{1}{4} 
    \iiint \d^2q_1\, \d^2q_3 \, \d^2q_4\,
    \widetilde{Z}_{N_1}^{M_1}(\bq_1+\bq_3/2)
    \widetilde{Z}_{N_2}^{M_2}(\bq_1+\bq_3/2)
    \widetilde{Z}_{N_1}^{M_1*}(\bq_1)
    \widetilde{Z}_{N_2}^{M_2*}(\bq_1)
    \widetilde{Z}_{N}^{M}(\bq_4)
    \widetilde{Z}_{N}^{M*}(\bq_3+\bq_4)
    \nn&\times
    \int_{|Q|\leq 2R} \d^2Q \, 
    e^{-\frac{3\gamma k}{2z}Q^2}
    \underbrace{
    \sum_{n_4'm_4'}
    \widetilde{Z}_{n_4'}^{m_4'}(2\bq_4-4\bq_1)
    Z_{n_4'}^{m_4'} \left(\frac{\bQ}{2R}\right)}_{\pi e^{2\pi i \left(\frac{1}{2R}\bar{\mathbf{Q}}\right) \cdot (4\bq_1-2\bq_4)}}
    \nn&\times
    \int\d^2s_3\,
    e^{2\pi i \bs_3\cdot 2\bq_3}
    \sum_{n_5'm_5'}
    G_{n_5'}^{m_5'}(\gamma)
    Z_{n_5'}^{m_5'}(\bs_3)
    \underbrace{
    \sum_{n_5m_5} \widetilde{Z}_{n_5}^{m_5}\left(-\frac{2kR}{z}\bQ\right)Z_{n_5}^{m_5}(\bs_3)}_{\pi e^{2\pi i \left(\frac{2kR}{z}\bar{\mathbf{Q}}\right) \cdot \mathbf{s}_3}}
    \nn&=
    \left(\frac{4k}{\pi z}\right)^2 \frac{\pi^2}{4} 
    \iiint \d^2q_1\, \d^2q_3 \, \d^2q_4\,
    \widetilde{Z}_{N_1}^{M_1}(\bq_1+\bq_3/2)
    \widetilde{Z}_{N_2}^{M_2}(\bq_1+\bq_3/2)
    \widetilde{Z}_{N_1}^{M_1*}(\bq_1)
    \widetilde{Z}_{N_2}^{M_2*}(\bq_1)
    \widetilde{Z}_{N}^{M}(\bq_4)
    \widetilde{Z}_{N}^{M*}(\bq_3+\bq_4)
    \nn&\times
    \int \d^2Q \, Z_0^0\left(\frac{\bQ}{2R}\right)
    e^{-\frac{3\gamma k}{2z}Q^2}
    e^{2\pi i \left(\frac{1}{2R}\bar{\mathbf{Q}}\right) \cdot (4\bq_1-2\bq_4)}
    \int\d^2s_3\,
    e^{2\pi i \bs_3\cdot 2\bq_3}
    \sum_{n_5'm_5'}
    G_{n_5'}^{m_5'}(\gamma)
    Z_{n_5'}^{m_5'}(\bs_3)
    e^{2\pi i \left(\frac{2kR}{z}\bar{\mathbf{Q}}\right) \cdot \mathbf{s}_3},
\end{align}
where $\bar{\bQ}=(Q,-\phi)$ is the vector $\bQ$ reflected around the x-axis, and we substituted $\mathcal{G}_{n_4'n_5}^{m_4'm_5} $ with \eqref{c18}. 
\paragraph{Final Discrete Tensor Collapse:}
In the Fraunhofer regime one can
drop the factors $e^{-\frac{3\gamma k}{2z}Q^2}$ and $e^{2\pi i \left(\frac{2kR}{z}\bar{\mathbf{Q}}\right) \cdot \mathbf{s}_3}$ and get
\begin{align}
&P_{N}^{M}\Big(\widetilde Z_{N_1}^{M_1}, \widetilde Z_{N_2}^{M_2}\Big) =
    \left(\frac{4kR}{z}\right)^2  
    \iiint \d^2q_1\, \d^2q_3 \, \d^2q_4\,
    \widetilde{Z}_{N_1}^{M_1}(\bq_1+\bq_3/2)
    \widetilde{Z}_{N_2}^{M_2}(\bq_1+\bq_3/2)
    \widetilde{Z}_{N_1}^{M_1*}(\bq_1)
    \widetilde{Z}_{N_2}^{M_2*}(\bq_1)
    \nn&\qquad\qquad\qquad\qquad\qquad \qquad\qquad\times
    \widetilde{Z}_{N}^{M}(\bq_4)
    \widetilde{Z}_{N}^{M*}(\bq_3+\bq_4)
    \widetilde{Z}_0^0(4\bq_1-2\bq_4)
    \sum_{n_5'm_5'}
    G_{n_5'}^{m_5'}(\gamma)
    \widetilde{Z}_{n_5'}^{m_5'}(2\bq_3)
    \nn&=
    \left(\frac{kR}{4z}\right)^2  
    \int \d^2u\, 
    \widetilde{Z}_{N_1}^{M_1}(\bu)
    \widetilde{Z}_{N_2}^{M_2}(\bu)
    \widetilde{Z}_{N}^{M*}(2\bu)
    \int\d^2u_1 \,
    \widetilde{Z}_{N_1}^{M_1*}(\bu_1)
    \widetilde{Z}_{N_2}^{M_2*}(\bu_1)
    \widetilde{Z}_{N}^{M}(2\bu_1)
    \sum_{n_5'm_5'}
    G_{n_5'}^{m_5'}(\gamma)
    \widetilde{Z}_{n_5'}^{m_5'}(2(\bu-\bu_1))
\end{align}
Note that the no-turbulence limit recovers the expected result \eqref{eq:P-final-noturb} as $\sum_{n_5'm_5'}G_{n_5'}^{m_5'}(\gamma) \widetilde{Z}_{n_5'}^{m_5'}(2(\bu-\bu_1))\to \sum_{n_5'm_5'}Z_{n_5'}^{m_5'}(0) \widetilde{Z}_{n_5'}^{m_5'}(2(\bu-\bu_1))\propto \pi $.
By applying the addition and linearization formulas \eqref{F_A_Z_proj} and \eqref{eq:F_kintner_janssen} the multi-dimensional integrals completely decouple into a discrete structure:
\begin{align}
P_{N}^{M}\Big(z_{N_1}^{M_1}, z_{N_2}^{M_2}\Big) &=
    \frac{1}{2\pi}\left(\frac{4kR}{z}\right)^2  
    \sum_{\substack{n_1n_2n_3n_4n_5\\ m_1m_2m_3m_4m_5}} G_{n_5}^{m_5}(\gamma) A_{n_1n_2n_5}^{m_1m_2m_5}(-1)^{n_4+n_3}
    \Gamma_{N_1N_2n_3}^{M_1M_2m_3}
    \Gamma_{N_1N_2n_4}^{M_1M_2m_4*}
    \Gamma_{n_3Nn_1}^{-m_3Mm_1*}
    \Gamma_{n_4Nn_2}^{-m_4M-m_2}.
\end{align}
Noting that $m_1=-m_2=M-M_1-M_2$, $m_5=0$, $m_3=m_4=M_1+M_2$, hence, $(-1)^{n_4+n_3}=(-1)^{m_4+m_3}=1$ and defining $F_n=\sum_{n'}
\Gamma_{N_1N_2n'}^{M_1M_2m'} 
\Gamma_{n'Nn}^{-m'Mm*}$, we finally arrive to
\begin{align}
P_{N}^{M}\Big(z_{N_1}^{M_1}, z_{N_2}^{M_2}\Big) &=
    \sum_{n_1n_2} \left[\sum_{n_5}G_{n_5}^0(\gamma)A_{n_1n_2n_5}^{m_1,-m_10}\right]F_{n_1}F_{n_2}^*.
\end{align}

\end{document}